\documentstyle[aps,version2,preprint,epsfig,rotating]{revtex}
\begin{document}
\draft
\preprint{OCIP/C 97-6}
%\preprint{September 1997}
\begin{title}
SINGLE $W$ BOSON PRODUCTION \\
IN HIGH ENERGY $e\gamma$ COLLISIONS
\end{title}
\author{Michael A. Doncheski$^1$ Stephen Godfrey$^2$  \\
and K. Andrew Peterson$^3$}
\begin{instit}
$^1$Department of Physics, Pennsylvania State University, Mont Alto, 
PA 17237 USA \\
$^2$Ottawa-Carleton Institute for Physics \\
Department of Physics, Carleton University, Ottawa CANADA, K1S 5B6 \\
$^3$Department of Physics and Physical Oceanography \\
Memorial University of Newfoundland, St. John's, NF, Canada. A1B 3X7 
\end{instit}
\begin{abstract}
We studied single $W$ boson production in high energy $e\gamma$
collisions and the sensitivity of various observables to the $WW\gamma$
gauge boson coupling by evaluating the helicity amplitudes of
all  Feynman diagrams which contribute to the final state being 
studied, including the $W$ decay to final state fermions.
We examined $W$ production
at 500 GeV and 1 TeV $e^+e^-$ colliders, comparing results for photon
spectra obtained from  a
backscattered laser and from beamstrahlung radiation.  Here we found that
the couplings could best be measured using the backscattered laser
photons with $|\delta\kappa_\gamma|\leq 0.07$ and
$|\lambda_\gamma| \leq 0.05$ at a 500 GeV collider and
$|\delta\kappa_\gamma|\leq 0.07$ and
$|\lambda_\gamma| \leq 0.02$ at a 1 TeV collider, all at 95\% C.L..
The measurement of $\kappa_\gamma$
is at the threshold of being able to measure loop contributions to
the trilinear gauge boson vertex.  
For completeness we include the limits achievable using
single $W$ production at a 200 GeV $e^+e^-$ collider
in the Weizacker-Williams approximation. This process can
measure $\kappa_\gamma$ to $\pm 0.15$ at 95\% C.L.
which is comparable to the $W$ pair production process. 

\end{abstract}
%\pacs{PACS numbers: 12.15.Ji, 14.80.Er, 12.50.Fk}

\narrowtext
\section{INTRODUCTION}
\label{sec:intro}

Despite the fact that the standard model of the electroweak interactions
\cite{sm} agrees
extraordinarily well with all existing measurements
\cite{smdata,lepdata} there is a widespread
conviction that it is nothing more than a low energy limit of a more
fundamental  theory \cite{bagger}.  An approach which is receiving 
considerable
attention is to represent new physics by additional terms in an
effective Lagrangian expansion and then to constrain the
coefficients of the effective Lagrangian by precision
experimental measurements \cite{bagger,falk91,non-linear,holdom}.
The bounds obtained on the coefficients
can then be related to possible theories of new physics.  For example,
this approach has been used by a number of authors to bound
dimension four operators which can contribute to the vacuum polarization
tensors of the massive gauge bosons via a global analysis of neutral
current data \cite{stu}.
These bounds have put severe constraints on technicolour
theories of dynamical symmetry breaking.  Similarly, the trilinear
gauge boson couplings have been described by effective
Lagrangians\cite{tgv,hagi87,gaemers}.
In one commonly used parametrization,  for on shell
photons,  the CP and P conserving
$\gamma WW$ vertex is parametrized in terms of two
parameters, $\kappa_\gamma$ and $\lambda_\gamma$ \cite{hagi87}.
Although bounds can be extracted from high precision low energy
measurements and measurements at the $Z^0$ pole \cite{fits}, 
there are ambiguities
and model dependencies in the results \cite{lowen}.  In contrast, gauge boson
production at colliders can measure the gauge boson couplings directly
and unambiguously.  The best current direct
measurement of these parameters come from the CDF and D0 
collaborations at the Fermilab $p\bar{p}$.  Their combined limits, 
varying one parameter at a time, are
 $|\delta \kappa_\gamma| < 0.9 $,  $|\lambda_\gamma| < 0.3$,
 $|\delta \kappa_Z| < 0.5 $, and  $|\lambda_Z| < 0.3$
 at 95\% C.L. \cite{tevatron}.
Putting tight constraints on the trilinear
gauge boson couplings is one of the primary motivations for the LEP200 upgrade
\cite{hagi87,zep87,lep200a,lep200b,l3,kane89}.
At the time of writing the 95 \% C.L. bounds from the LEP experiments 
\cite{lep200b} are  $-0.9 \stackrel{<}{\sim} \delta \kappa 
\stackrel{<}{\sim} 1.12$
and $ -0.78 \stackrel{<}{\sim} \lambda \stackrel{<}{\sim} 1.19$ 
where only one 
parameter is varied at a time and these results have assumed that
$\delta \kappa_\gamma=\delta\kappa_Z $ and $\lambda_\gamma =\lambda_Z $.
The L3 collaboration at CERN has also obtained the 95\% C.L. bounds 
from single $W$ production \cite{l3}; $-3.6 < \delta \kappa_\gamma < 1.5 $
and $ -3.6 < \lambda < 3.6$.
Precise direct measurement to the level of several percent will have to
wait until the era of the Large Hadron Collider \cite{kane89,ssc/lhc} 
and the NLC, a high energy $e^+e^-$ collider.

Recently, the idea of constructing $e\gamma$ and $\gamma\gamma$
colliders using either high energy photons from lasers backscattered
off of a high energy electron beam \cite{laser} or photons arising from
beamstrahlung radiation \cite{beam1,beam2,beam3}
has received serious attention.
The physics possibilities for $e\gamma$ colliders are the subject of a
growing literature \cite{egamma}. In particular, the properties of $W$ bosons,
including the $\gamma WW$ coupling, has been examined in a number of
recent publications \cite{egtow,yehudai}.  The LEP 
collaborations \cite{l3} are starting to obtain results for single $W$ 
production where the photon arises in t-channel exchange \cite{singlew}
although the 
results are still rather weak compared to those obtained by CDF and D0.

In this paper we reexamine the sensitivity to which the $\gamma WW$
vertex can be measured at $e\gamma$ colliders using
photon spectra produced from backscattered lasers and beamstrahling
radiation.
In contrast to other analyses \cite{yehudai} we include the
decay of the $W$ boson to final state fermions along with
contributions to the final state that do not proceed via an
intermediate $W$ boson.  At high energy this provides additional
information off the $W$ resonance through the interference of the various
diagrams.  In addition, we considered the various backgrounds that may 
obscure results using hadronic $W$ decay.

We focus on  $W$
production at 500 GeV and 1 TeV $e^+e^-$ colliders and compare the
sensitivities achievable using a backscattered laser photon spectrum
and a beamstrahlung photon spectrum.
For completeness we also include results obtained from
single $W$ production at a 200 GeV $e^+e^-$ collider using the 
Weizacker-Williams effective photon approximation as
a competing process to $W$ pair production.

In the next section  we write down the effective vertex and the resulting
Feynman rule. In Sec.~\ref{sec:results}
we present our calculation and results and in the final
section we give our conclusions.

\section{THE $WW\gamma$ EFFECTIVE VERTEX}
\label{sec:vertex}

Within the standard model the $WW\gamma$ vertex is uniquely determined
by $SU(2)_L \times U(1)$ gauge invariance so that a precise measurement
of the vertex poses a severe test of the gauge structure of the theory.
The most general $WW\gamma$ vertex, consistent with Lorentz invariance,
can be parametrized in terms of seven form factors when the
W bosons couple to essentially massless fermions which effectively results
in $\partial_\mu W^\mu=0$ \cite{hagi87,gaemers}.  For on shell photons,
electromagnetic gauge invariance further restricts the tensor structure
of the $WW\gamma$ vertex to allow only four free parameters, two of
which violate CP invariance.  Measurement of the neutron
electric dipole moment constrains the two CP violating parameters,
$\tilde{\kappa}_\gamma$ and $\tilde{\lambda}_\gamma$,
to values
too small to give rise to observable effects in the process we are
considering;
$|\tilde{\kappa}_\gamma|, |\tilde{\lambda}_\gamma|<
{\cal O} (10^{-4})$ \cite{marci86,boudje91}.
Therefore, the most general  Lorentz and CP invariant vertex compatible with
electromagnetic gauge invariance is commonly parametrized as \cite{hagi87}:
\begin{equation}
{\cal L}_{WW\gamma} =  - ie \left\{ {(W^\dagger_{\mu\nu}W^\mu A^\nu -
W^\dagger_\mu A_\nu W^{\mu\nu} )
+ \kappa_\gamma W^\dagger_\mu W_\nu F^{\mu\nu}
+ {{\lambda_\gamma}\over{M_W^2}} W^\dagger_{\lambda\mu}W^\mu_\nu F^{\nu\lambda}
}\right\}
\end{equation}
where $A^\mu$ and $W^\mu$ are the photon and $W^-$ fields,
$W_{\mu\nu}=\partial_\mu W_\nu - \partial_\nu W_\mu$
and $F_{\mu\nu}=\partial_\mu A_\nu - \partial_\nu A_\mu$
denote the $W$ and photon field strength tensors,
and $M_W$ is the $W$ boson mass.
Higher dimension operators correspond to momentum dependence in the form
factors.  The first term, referred to as the minimal coupling term,
simply reflects the charge of the $W$.
The Feynman rule for the $WW\gamma$ vertex resulting from eqn. (1) is
given by:
\begin{eqnarray}
 ie \{ &
g_{\alpha\beta}[(1-\hat{\lambda} \; k_- \cdot q ) {k_+}_\mu
-(1-\hat{\lambda} \; k_+ \cdot q ) {k_-}_\mu ] \nonumber \\
& - g_{\alpha\mu}[(1-\hat{\lambda} \; k_- \cdot q ) {k_+}_\beta
-(\kappa-\hat{\lambda} \; k_+ \cdot k_- ) q_\mu ] \nonumber \\
&- g_{\beta\mu}[(\kappa-\hat{\lambda} \; k_- \cdot k_+ ) q_\alpha
-(1-\hat{\lambda} \; k_+ \cdot q ) {k_-}_\alpha ] \nonumber \\
&+\hat{\lambda} ({k_+}_\mu {k_-}_\alpha q_\beta
-{k_-}_\mu q_\alpha {k_+}_\beta ) \}
\end{eqnarray}
with the notation and conventions given in Fig. 1 and where
$\hat{\lambda}=\lambda/M_W^2$.
At tree level the standard model predicts $\kappa_\gamma=1$ and
$\lambda_\gamma=0$.  Other parametrizations exist in the literature such
as the chiral Lagrangian expansion  and one can map the
parameters we use to those used in other approaches 
\cite{falk91,non-linear}.

\newpage

\centerline{\epsfig{file=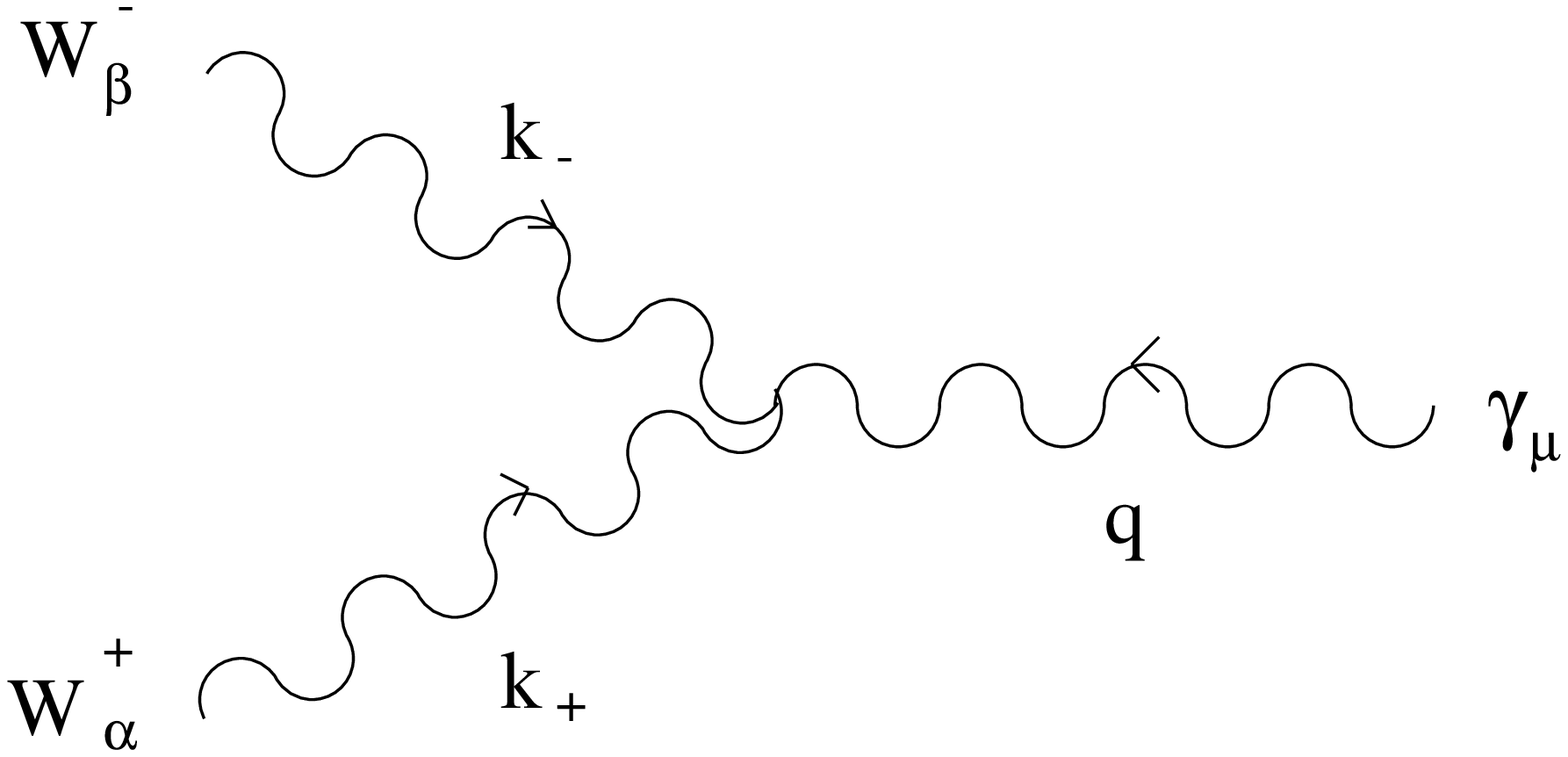,width=4.0cm,clip=}}

\noindent
{\small {\bf Figure 1:} Feynman diagram for the $WW\gamma$
vertex corresponding to the Lagrangian
and Feynman rule given in the text.}

Before proceeding it is useful to comment on what to expect for
$\kappa_\gamma$ and $\lambda_\gamma$.  From the chiral Lagrangian approach
one expects that
$\delta\kappa_\gamma \sim O(10^{-2})$ and  $\lambda_\gamma$ is suppressed by
an additional factor of 100 \cite{falk91}.  These order of magnitude estimates
are
confirmed by explicit calculation.  The contribution of the $t$-quark
results in $\delta \kappa \simeq 1.5\times 10^{-2}$ and $\lambda_V
\simeq +2.5\times 10^{-3}$ while a SM Higgs boson of mass 200 GeV
contributes $\delta\kappa \simeq 5\times 10^{-4}$ and $\lambda \simeq
4\times 10^{-5}$ \cite{couture-t}. Technicolour theories give $\delta\kappa_Z
=-0.023$
and $\delta\kappa_\gamma =0$ \cite{holdom} and supersymmetric theories
give $\delta\kappa_{max} \simeq 7\times 10^{-3}$ and
$\lambda_{max}\simeq 10^{-3}$ \cite{couture-s}.

What one gleans from these numbers is
that a deviation of more than a couple of percent would be difficult
to accomodate in the SM and in general, contributions via loop
corrections  typically contribute no more than a couple of percent.
Although it might be possible to find models that give slightly larger
contributions, for example models with $Z-Z'$ mixing \cite{frere} or models
with many particles
which contribute coherently to loop contributions \cite{golden}, if
deviations much larger than several percent are observed this would signal
something very radical such as composite gauge bosons \cite{suzuki}.
Since we know
of no convincing models of this sort, to probe for new physics via
anomalous trilinear gauge boson couplings one must be able to measure
the vertex to the level of a few percent.

Deviations from the standard model ($a=\delta \kappa= \kappa-1$,
$\lambda$) lead to amplitudes which grow with energy
and therefore violate unitarity at high energy \cite{unitarity,baur}.
One method of avoiding violation of the unitarity bound is to include a
momentum dependence in the form factors, $a(q^2_W, \bar{q}^2_W,
q_\gamma^2 =0)$,  so that the deviations vanish
when either $|q_W^2|$ or $|\bar{q}_W^2|$, the absolute
square of the four momentum of the vector bosons, becomes large
\cite{baur}.  We
therefore  include the form factors
\begin{equation}
a(q_W^2, \bar{q}_W^2, 0) = a_0 [(1+|q_W^2|/\Lambda^2)
(1+|\bar{q}_W^2|/\Lambda^2)]^{-n}
\end{equation}
where $\Lambda$ represents
the scale at which new physics becomes important and $n$ is chosen
as the minimum value compatible with unitarity.
We take $n=1$ and $\Lambda=1$ TeV in our numerical results.  We find
that lowering this scale effects our conclusions slightly although 
in this case the new physics should show up elsewhere. Increasing the
scale has only a small effect on our results.

\section{CALCULATION AND RESULTS}
\label{sec:results}

The Feynman diagrams contributing to the process $e^-\gamma \to \nu
f\bar{f}$ are given in Fig. 2.  The
$WW\gamma$ vertex we are studying contributes via diagram 2b.  To
preserve electromagnetic gauge invariance and to properly take into
account the background processes our calculation includes all the
diagrams of Fig 2 for arbitrary values of $\kappa_\gamma$ and
$\lambda_\gamma$.  To obtain the cross sections and distributions we
used the CALKUL helicity amplitude technique \cite{calkul} which for
completeness we
summarize in Appendix A along with the amplitudes corresponding to Fig. 2.
Monte Carlo integration techniques are then used to
perform the phase space integrals \cite{monte}.  We treat the photon
distributions
as structure functions, $f_{\gamma/e}(x)$ and integrate them with the
$e\gamma$ cross  sections to obtain our results:
\begin{equation}
\sigma = \int f_{\gamma/e}(x) \sigma (e\gamma\to \nu f\bar{f})\; dx
\end{equation}
where $x$ is the fraction of the original electron energy carried by
the photon.
For completeness and for the convenience of the interested reader we
include the various photon distributions in Appendix B.
For our numerical results we take $\alpha(M_Z)=1/128$,
$M_W=80.22$ GeV, $\Gamma_W=2.0$ GeV, $\sin^2\theta_w=0.23$.

\centerline{\epsfig{file=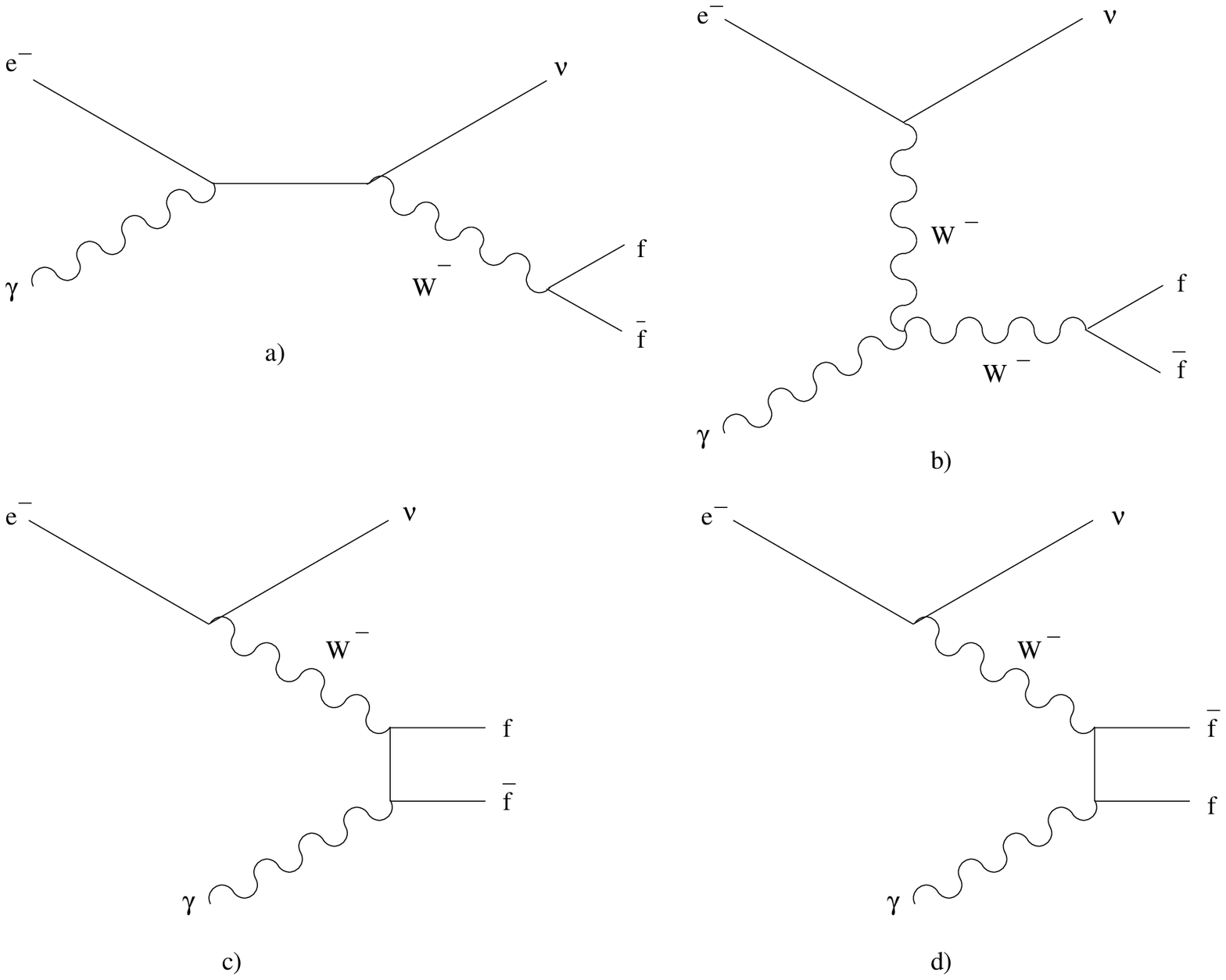,width=5.0cm,clip=}}

\noindent
{\small {\bf Figure 2:} The Feynman diagrams contributing to the process $e\gamma \to
\nu q\bar{q}$. For the process $e\gamma \to \nu_e \mu \bar{\nu}_\mu$
diagram (c) does not contribute and the quark charge in diagram (d)
should be replaced by the $\mu$ charge.}

The signal we are studying consists of either, (i) for leptonic $W$ decay,
a high transverse momentum lepton ($p_T$) and
large missing transverse momentum ($\not{p}_T$) due to the neutrinos
from the initial electron beam and from the $W$ decay, or (ii) for hadronic
$W$
decay, two hadronic jets and large missing transverse momentum ($\not{p}_T$)
due to
the neutrino from the initial electron
beam.  In both cases, we impose the acceptance cut 
that visible particles in the final state
be at least $10^o$ from the beam direction.
Our conclusions are not sensitive to the exact value of this cut.  We
also impose a cut on the minimum $\not{p}_T$; for $\sqrt{s}=200$~GeV 
we used  $\not{p}_T > 5$~GeV and for $\sqrt{s}=500$~GeV and 1~TeV
we used $\not{p}_T > 10$~GeV.
We do not include fragmentation and hadronization effects for the
hadronic modes and identify the hadron jet momenta with that of the
quarks.  The signals we consider are therefore
\begin{eqnarray}
& e^- + \gamma \to \mu^- + \not{p} \\
& e^- + \gamma \to j  + j  + \not{p}
\end{eqnarray}
We also examined the reaction $e^- \gamma \to e \bar{\nu} \nu$ which
includes the process $e^-\gamma \to e^- Z^0 \to e^- \nu \bar{\nu}$
in addition to the diagrams of figure 2.  However, once kinematic cuts
are imposed that eliminate the uninteresting contributions from
$e\gamma \to e^- Z$ we are left with results comparable to the $\mu$
mode.

\subsection{Backgrounds}

Before proceeding we discuss potential backgrounds to the processes we 
are studying \cite{background}.  They can be divided into 3 categories:
\begin{enumerate}
\item Direct
\begin{eqnarray}
e^- \gamma & \to & e^- Z^0 \to e^- q\bar{q} \\
e^-e^+ & \to &\gamma Z^0  \to \gamma q\bar{q}
\end{eqnarray}
where in the first case the outgoing $e^-$ is not observed while in 
the second case the outgoing $\gamma$ is not observed. 

\item Once Resolved.  These involve $\gamma q$ or $\gamma g$ 
collisions where 
the $q$ or $g$ is treated as a parton in the photon.  The subprocesses 
are: 
$\hat{\sigma}(\gamma q \to q g)$, 
$\hat{\sigma}(\gamma \bar{q} \to \bar{q} g)$, 
and $\hat{\sigma}(\gamma g \to q \bar{q})$.
\item Doubly Resolved.  Here, $q$ or $g$ partons come from both the 
$\gamma$ and a Weizacker-Williams photon coming from the electron.
The subprocesses are: 
$\hat{\sigma}(g g \to q\bar{q})$, 
$\hat{\sigma}(gg \to gg)$, 
$\hat{\sigma}(q g \to q g)$, 
$\hat{\sigma}(q\bar{q} \to gg)$, 
$\hat{\sigma}(q\bar{q} \to q\bar{q})$, 
$\hat{\sigma}(q\bar{q} \to q'\bar{q}')$, 
$\hat{\sigma}(q q \to qq )$, 
$\hat{\sigma}(q\bar{q}' \to q\bar{q}')$, 
and $\hat{\sigma}(q q' \to q q')$.

\end{enumerate}

The $e^- e^+ \to Z^0 \gamma$ is easily removed by imposing the 
constraint that the photon not be observed while at the same time 
there is significant missing $p_T$ in the event.  Since the photon 
recoils against the $Z$ these two contradictory requirements eliminate 
this background.  We show the $jet-jet$ 
invariant mass for the signal and remaining backgrounds for the 
backscattered laser case with $\sqrt{s}=500$~GeV in Fig. 3. These 
results include
the detector acceptance cuts of $10^o <\theta_{e-jet} < 170^o$ and
$p_T(jet)>5$~GeV.  The resolved photon backgrounds are substantial but 
they can be eliminated by requiring a minimum missing $p_T$.  This is 
a simple consequence of the fact that we require both final state 
particles to be observed while the initial state partons are essentially 
collinear with the beam direction.  The only remaining background
is the direct background 
$e^-\gamma \to e Z$ with the electron escaping down the beam pipe. We 
eliminate this background by 
imposing a tight cut on the invariant mass of the outgoing jets so 
only jets contributing to the $W$ peak remain. This
eliminates the bulk of the background from the large $Z$ peak.  For the 
case of dijet pairs with invariant mass 
off the $W$ resonance we can eliminate the $e^-\gamma \to e Z$ 
background by imposing a much larger cut on the missing $p_T$.  Since 
we eliminate events when the electron is observed no events will 
remain with missing $p_T$ greater than the maximum $p_{T_e}$ for an 
electron lost down the beam pipe ($\sim 40$~GeV for $\sqrt{s}=500$~GeV).

\centerline{
\begin{turn}{-90}
\epsfig{file=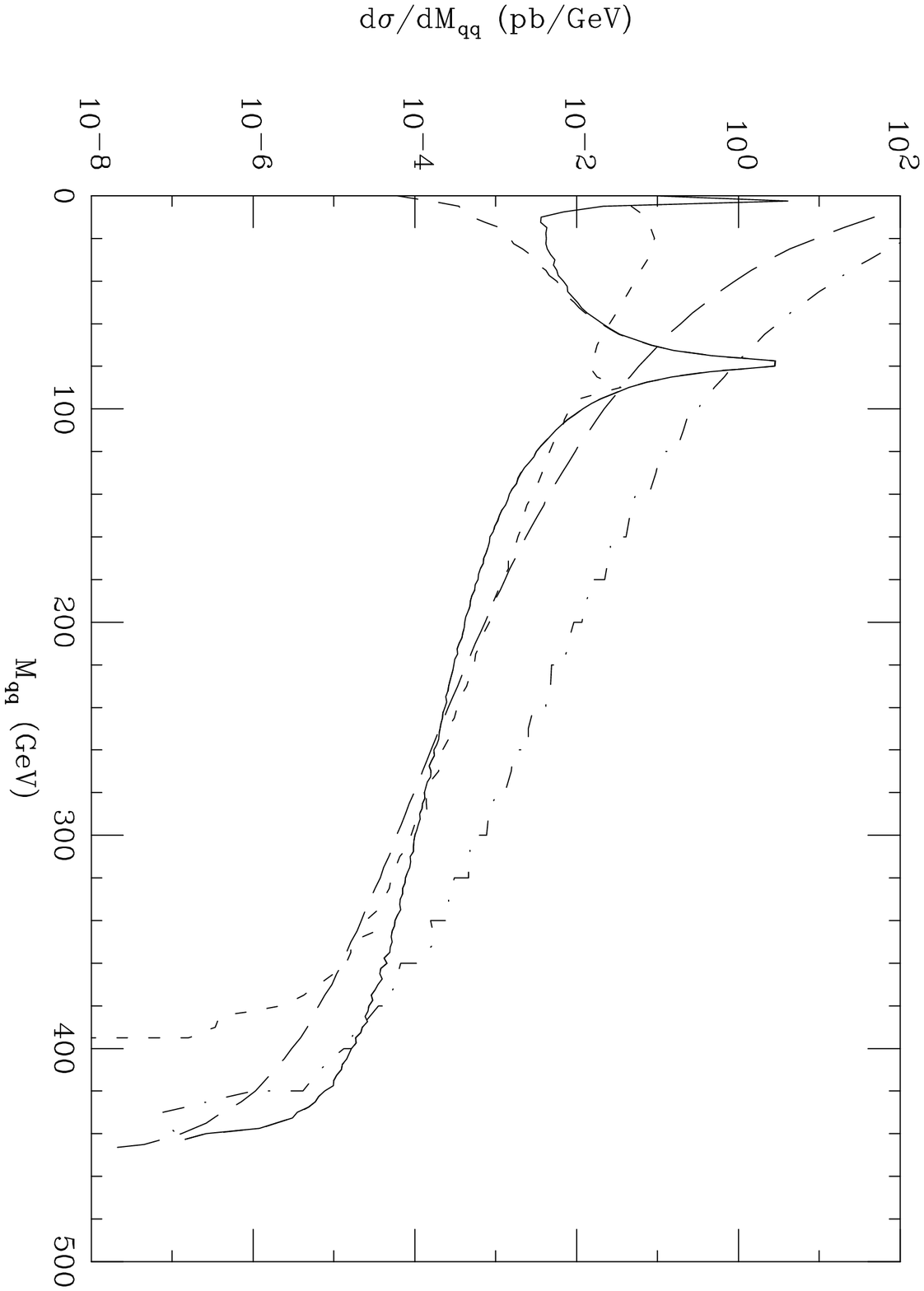,width=5.5cm,clip=}
\end{turn}
}

\noindent
{\small {\bf Figure 3:}
Dijet invariant mass for the signal and backgrounds.  The 
solid line is the signal, while the dashed line is the signal with the 
detector resolution $p_{_T}$ cut.  The short dashed curve is the 
$e \gamma \rightarrow e Z$ background, the long dashed curve is the 
total of the singly resolved backgrounds and the dotdashed curve is the 
total of the doubly resolved backgrounds.}

\subsection{Results}

In the following subsections we study single $W$ production for
$\sqrt{s}=200$~GeV, 500~GeV, and 1~TeV.
In doing so we examined a number of kinematic
distributions,  asymmetries, and ratios of observables  for their
sensitivity to anomalous gauge boson couplings.  Since we will only
present results for the observables most sensitive to
the gauge boson couplings, at this point we
list all observables that we studied. For the leptonic $W$ decay modes we
considered;  $d\sigma/dp_{T\mu}$, $d\sigma/dE_{\mu}$,
$d\sigma/d\cos\theta_{e\mu}$, where $\theta_{e\mu}$ is the angle
between the incoming electron and the outgoing muon,
$A_{FB}=(\sigma_F-\sigma_B)/(\sigma_F+\sigma_B)$,
$R_{IO}=\sigma_I/\sigma$ where $\sigma_I$ is the cross section for
$|\cos\theta_{e\mu}|<0.4$ and $\sigma$ is the total cross section with
the kinematic cuts given above.  When backscattered laser photons are
used we can form the ratio $R_{13}=\sigma_1/\sigma_3$ where
$\sigma_1$ is the cross section for the mainly helicity 1/2 amplitudes
which result when the incident photons are left handed and $\sigma_3$
is the cross section for the mainly helicity 3/2 amplitudes when the
incident photons are right handed\footnote{Note that the high energy photon
beam
has opposite polarization to that of the laser.}.
For the hadronic $W$ decay modes we
reconstructed the $W$ boson 4-momentum from the hadronic jets'
4-momentum, imposing the kinematic cut of 75 GeV
$<M_{q\bar{q}}=\sqrt{(p_q +p_{\bar{q}})^2}<$ 85 GeV.
Including the nonresonant diagrams of fig (2c) and (2d) and reconstructing
the $W$ boson in this manner gives different results than from simply
studying the cross sections to $W$ bosons.  Finally, in some cases,
which we will describe in further detail below, we studied the effect
of anomalous couplings on the cross section of the hadronic modes {\it
off} the $W$ pole (i.e. $M_{q\bar{q}} \neq M_W$).
In general, deviations of the gauge boson couplings had
a very substantial effect on the cross section off the $W$ resonance
although, because of the reduced cross section, the statistical
significance is not really enhanced.  This does, however,
point out the importance of considering the process that is actually
measured, not just a theorist's idealization.

\subsection{$\sqrt{s}=200$ GeV: LEP2}

This process has been examined in detail by a number of authors and 
the various LEP collaborations have presented preliminary results.  
As our results are consistent with other work, we include 
the limits that are achievable at LEP2 mainly for completeness.
At LEP 200 the possibility of $e\gamma$ collisions is only possible
via t-channel photon exchange between the electron and positron beams
\cite{singlew}.  We
therefore examine single $W$ production in $e^+e^-$ collisions by
folding the cross sections for $e\gamma\to q\bar{q}\nu$ and
$e\gamma\to \mu\bar{\nu}\nu$ with the Weizacker-Williams photon
distribution.

We find that distributions utilizing $W$'s reconstructed from the hadronic
modes are most sensitive to the $WW\gamma$ couplings.  This is primarily
due to the increased statistics of the hadronic modes over the
leptonic modes since the bounds obtained at 200 GeV are limited by
statistics; $\sigma_\mu = 0.059$~pb vs $\sigma_W=0.28$~pb (with the cut on
$M_{q\bar{q}}$).
To obtain sensitivity bounds we used the total cross
section for the reconstructed $W$ boson.
The 95\% C.L. limits are given in Fig. 4.
To obtain this curve we calculated the observable for a
given value of $\kappa_\gamma$ and $\lambda_\gamma$ and determined at
what level the value, if measured, would be compatible with the
standard model prediction. 
The 95\% C.L. limits obtained by varying one parameter at a time, for
$L=500$~pb$^{-1}$,
are $0.85 < \kappa_\gamma < 1.13$
and $-0.63 < \lambda_\gamma< 0.61$ from $\sigma(e^+e^-\to e^+ W^-)$.
The sensitivity on $\kappa_\gamma$ is comparable to that obtained from the
$W$-pair production process but because it offers a means of measuring the
$WW\gamma$ vertex independently of the $WWZ$ vertex, it is 
an important complement to the $W$-pair production process.

\vskip 0.2cm
\centerline{\epsfig{file=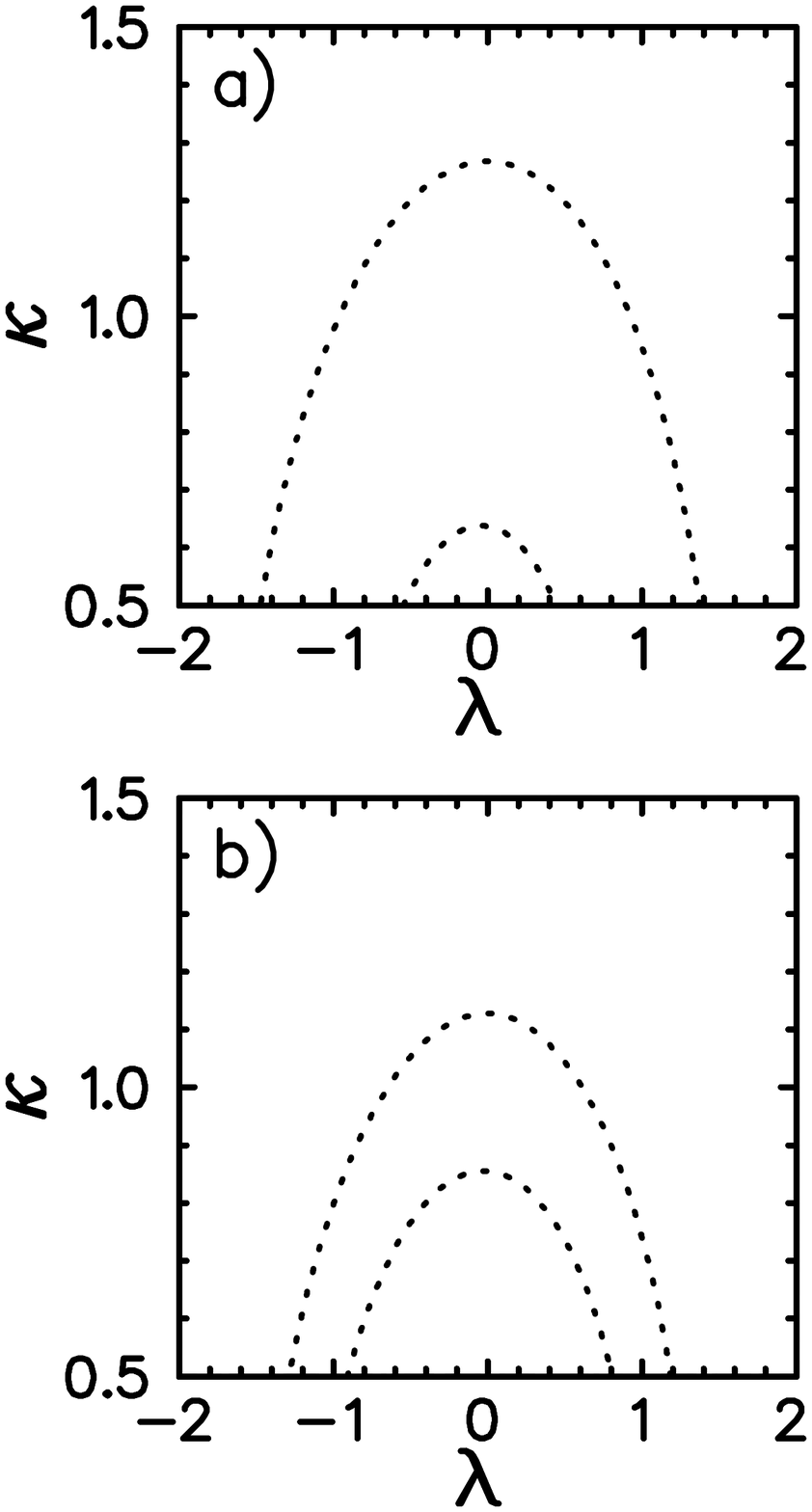,width=4.0cm,clip=}}

\noindent
{\small {\bf Figure 4:}
The achievable bounds on $\kappa_\gamma$ and $\lambda_\gamma$
at 95\% C.L. for a $\sqrt{s}=200$~GeV $e^+e^-$ collider
using the total cross section for the hadronic decay mode of the $W$ boson.
Fig. (a) is for $L=100$~pb$^{-1}$ and fig. (b) is for
$L=500$~pb$^{-1}$. }

\subsection{The Next Linear Collider; $\sqrt{s}=500$ GeV and
$\sqrt{s}=1000$ GeV}

For the NLC \cite{nlc}
we consider two possibilities for the photon spectrum; that
arising from backscattering a laser from one of the original electron beams and
beamstralung, which is the radiation which arises when two intense
beams of electrons pass through one another.  For the results using
the beamstrahlung photon spectrum we  concentrate on the beam spectrum
resulting from
the G set of parameters of Ref. \cite{palmer}.  We will discuss the effect
of different beam parameters on our results.  We included the
Weizacker-Williams contributions in our beamstrahlung results.

The NLC is envisaged as a very high luminosity collider so that the
number of events per unit of R, the QED point cross section, which is
an s-channel process and goes like $1/s$, remains reasonable.  The
integrated luminosities for a Snowmass year ($10^7$ sec) are expected
to be $\sim 60$ fb$^{-1}$ for a $\sqrt{s}=500$ GeV collider and $\sim 200$
fb$^{-1}$ for a $\sqrt{s}=1$~TeV collider.  Typical cross sections for
the process $e\gamma\to \mu \bar{\nu}_\mu \nu_e$ and
$e\gamma\to W \nu \to q \bar{q} \nu$ at $\sqrt{s}=500$ GeV
are 3.2 pb and 16.6 pb respectively for the backscattered laser mode
and 2.1 pb and 10.8 pb respectively for the beamstrahlung mode, leading to
$\sim 10^6$ events per year.
At $\sqrt{s}=1$~TeV
$\sigma(e\gamma\to \mu \bar{\nu}_\mu \nu_e)=4.0$~pb and
$\sigma(e\gamma\to W \nu \to q \bar{q} \nu)=19$~pb for the
backscattered laser mode and
$\sigma(e\gamma\to \mu \bar{\nu}_\mu \nu_e)=6.0$~pb and
$\sigma(e\gamma\to W \nu \to q \bar{q} \nu)=31$~pb for the
beamstrahlung mode  leading to $\sim 6\times 10^6$ events/year.
Thus, except for certain regions of phase space,
the errors are not limited by
statistics, but rather by systematic errors.
Factors entering into systematic errors include an accurate knowledge of
the total luminosity, particle misidentification,  triggering and
detector efficiencies, uncertainty in the
size of backgrounds, calorimetric accuracy etc.  Estimating systematic
errors requires detailed Monte Carlo studies which we do not attempt.
For cross sections we
assume a systematic error of 5\% and for asymmetries and ratios, where
some of the systematic errors cancel, we assume a systematic error of
3\% \cite{barklow}.  We consider the effects of reducing these systematic
errors on the achievable sensitivities.
In our results we combine in quadrature the statistical errors
based on the integrated luminosities given above with the systematic
errors:
\begin{equation}
\delta^2 = \delta^2_{stat} + \delta^2_{sys}.
\end{equation}

We begin with $\sqrt{s}=500$~GeV.
The total cross sections and the angular distributions of the outgoing muon
and reconstructed $W$ (from the hadronic decay mode)
are sensitive to anomalous couplings.  We plot
the distributions for both the backscattered photon case and the
beamstrahlung case in Fig. 5.  At higher energies we can
obtain additional information, especially for $\lambda_\gamma$,
from the $p_T$ spectrum of the outgoing
lepton or the reconstructed $W$.  We show these spectra in Fig. 6.
Finally, as already pointed out, the invariant mass distribution,
shown in Fig. 7, of the
$q\bar{q}$ pair above the $W$ mass also provides useful information.  If,
for example, we integrate the $M_{q\bar{q}}$ spectrum from 100 GeV up,
we obtain a cross section of 0.25 pb for the backscattered laser mode
which offers
considerable statistics.  For $M_{q\bar{q}}>300$~GeV, $\sigma= 0.006$~pb
which yields $\sim 400$ events/year.  More importantly, this high
$M_{q\bar{q}}$
region shows a higher sensitivity to anomalous couplings than the $M_W$ pole
region.

\vskip 0.2cm
\centerline{\epsfig{file=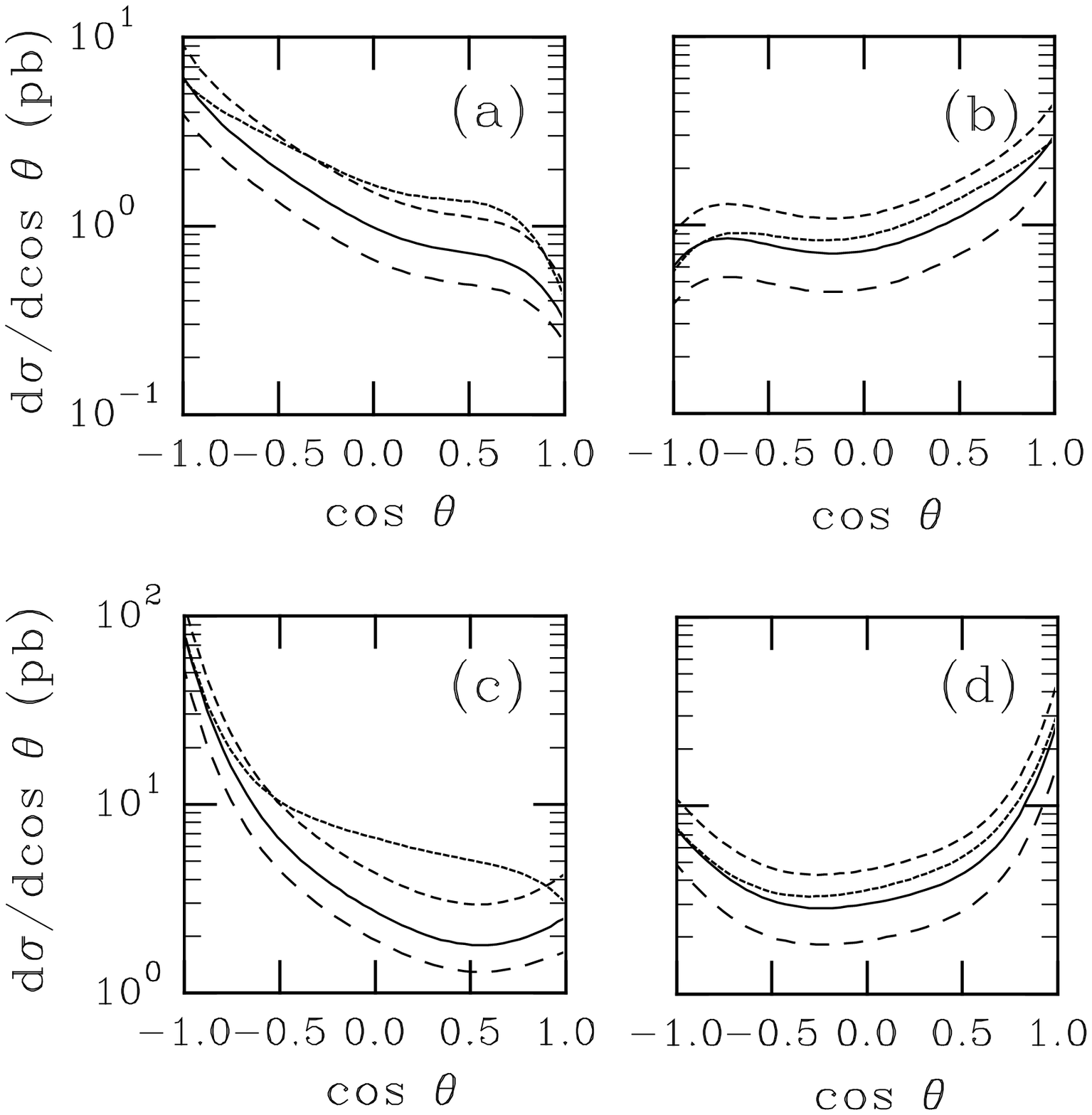,width=7.0cm,clip=}}

\noindent
{\small {\bf Figure 5:}
The angular distributions of the outgoing muon and
reconstructed $W$ boson
relative to the incoming electron for $\sqrt{s}=500$~GeV. (a) For a
muon with the backscattered laser photon spectrum, (b) for a muon
with the beamstrahlung photon spectrum, (c) for a reconstructed
$W$ boson with the backscattered laser photon spectrum, and (d) for a
$W$ boson with the beamstrahlung photon spectrum.  In all cases
the solid line is the standard model
prediction, the long-dashed line is for $\kappa_\gamma=0.6$,
$\lambda_\gamma=0$,  the short-dashed line is for $\kappa_\gamma=1.4$,
$\lambda_\gamma=0$,  and the dotted line is for $\kappa_\gamma=1$,
$\lambda_\gamma=0.4$.}

\centerline{\epsfig{file=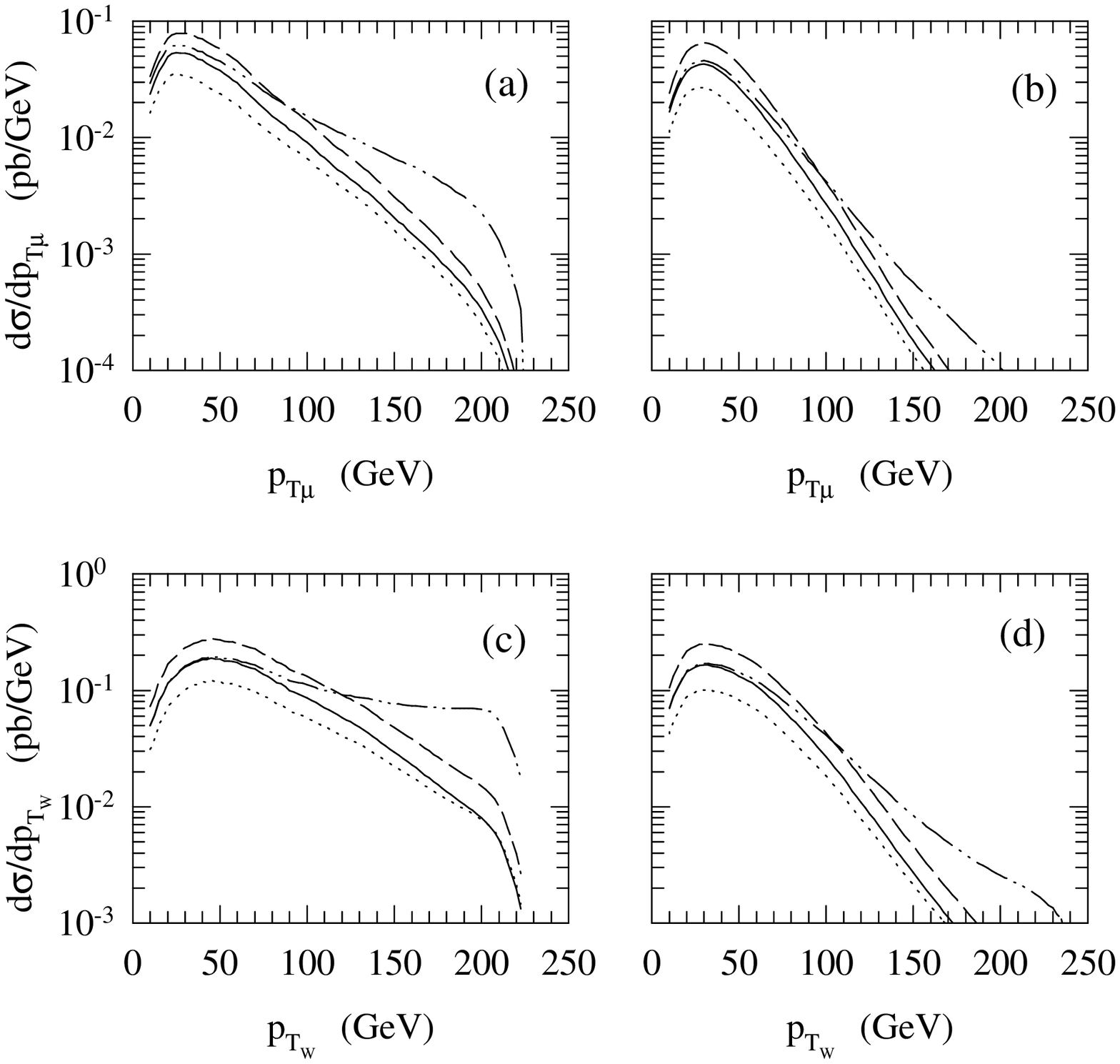,width=7.0cm,clip=}}

\noindent
{\small {\bf Figure 6:}
The $p_T$ distributions of the outgoing muon and reconstructed
$W$ boson for $\sqrt{s}=500$~GeV.
(a) For a muon with the backscattered laser photon spectrum, 
(b) for a muon with the beamstrahlung photon spectrum, 
(c) for a reconstructed 
$W$ boson with the backscattered laser photon spectrum, and 
(d) for a $W$ boson with the beamstrahlung photon spectrum.  
In all cases the solid line is the standard model prediction, 
the dotted is for $\kappa_\gamma=0.6$ and $\lambda_\gamma=0$,  
the long-dashed line is for $\kappa_\gamma=1.4$, and $\lambda_\gamma=0$,  
and the dot-dot-dashed line is for $\kappa_\gamma=1$, $\lambda_\gamma=0.4$.}

\centerline{\epsfig{file=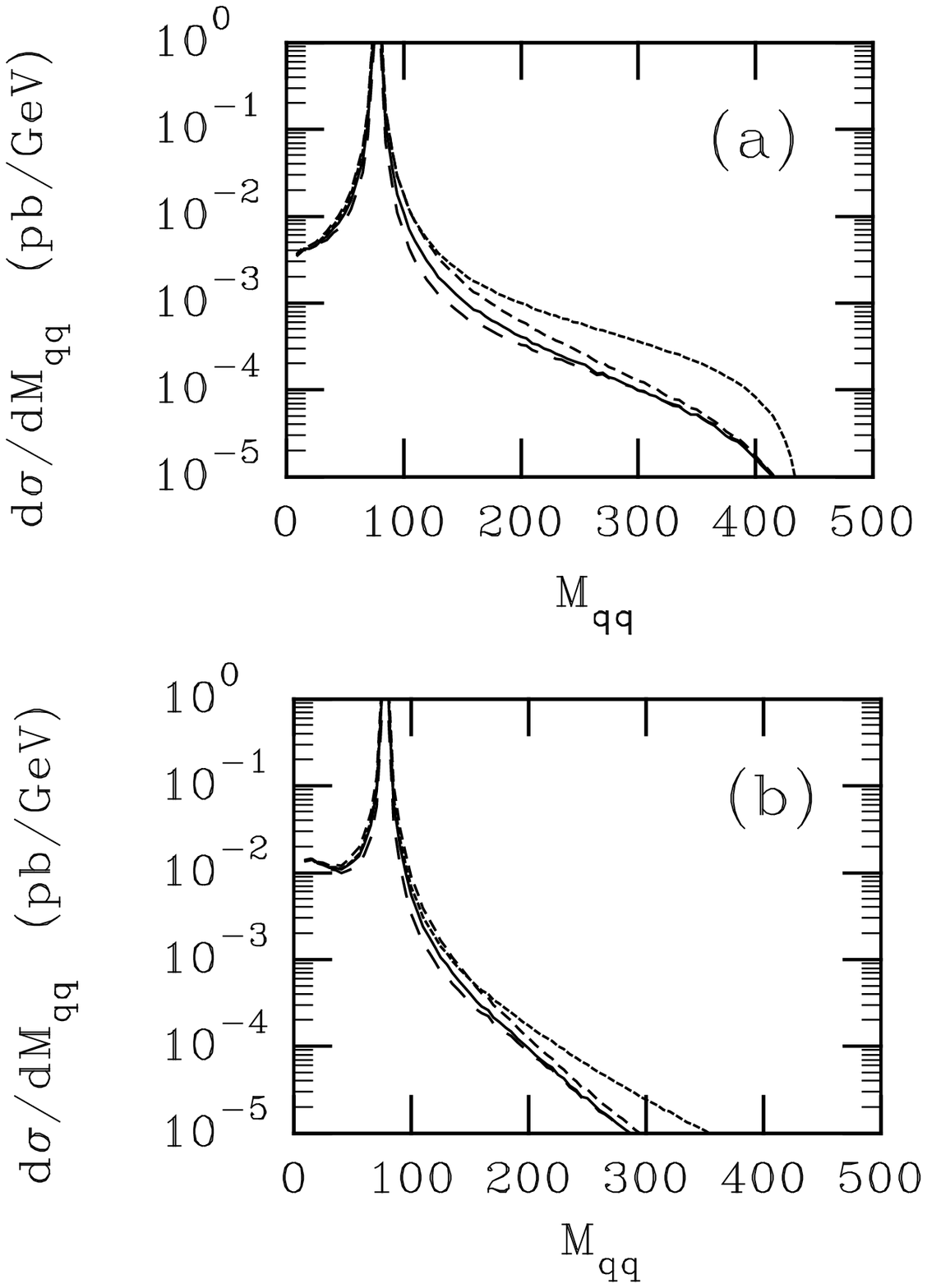,width=7.0cm,clip=}}

\noindent
{\small {\bf Figure 7:}
The hadron jet invariant mass ($M_{q\bar{q}}$) distribution
for $\sqrt{s}=500$~GeV. (a) For
the backscattered laser photon spectrum and (b) for
the beamstrahlung photon spectrum. In both cases
the solid line is the standard model
prediction, the long-dashed line is for $\kappa_\gamma=0.6$,
$\lambda_\gamma=0$,  the short-dashed line is for $\kappa_\gamma=1.4$,
$\lambda_\gamma=0$,  and the dotted line is for $\kappa_\gamma=1$,
$\lambda_\gamma=0.4$.}

To quantify the observations of the above paragraph we consider the
cross sections, the angular distributions, and the $p_T$ distributions
for the muon  and hadronic modes. 
For the angular distributions we used four
equal bins and for the $p_T$ distributions we used the 4 $p_T$ bins;
$0-100$~GeV, 100-150~GeV, 150-200~GeV, and 200-250~GeV.
The bounds obtained for these observables are shown in Fig. 8 based
on systematic errors of 5\%.  Bounds obtained by varying one parameter
at a time are summarized in Table I.  
In general the  limits on $\kappa_\gamma$
obtained using the beamstrahlung spectrum are comparable to
those obtained from the backscattered photon spectrum 
with the exception that we do not show results 
using angular distributions in the 
beamstruhlung mode because of the ambiguity in deciding the angle due 
to the poor charge determination of $W$'s decaying hadronically.
If the muon charge is identified in the beamstrahlung case, 
the angular distribution method  could be used as described, although 
we do not show these results in Fig 8 and 11.
The bounds on $\lambda_\gamma$ are tighter using the backscattered laser
photons which reflects the harder photon spectrum in this case to which
$\lambda_\gamma$ is more sensitive.
We find that $\kappa_\gamma$ can be measured to within 7\% and
$\lambda_\gamma$ to within $\pm 0.05$ at 95\% C.L., using the
backscattered laser approach, which is approaching
the sensitivity required to observe the contributions of new physics
at the level of loop corrections.  One can also obtain limits using 
the high invariant mass region.  A priori one might expect single 
$W$ production to be overwhelmed by the $e\gamma \to Z e$ background.  
However, as pointed out in a previous section,  
this background can be eliminated 
in the kinematic region of interest with an appropriate kinematic cut,
$\not{p}_T >40$~GeV, which has little effect on the signal.  We find 
that for $M_{q\bar{q}}>100$~GeV we obtain the 95\% C.L. sensitivity of 
$-0.09 <\delta \kappa_\gamma < 0.08$ 
($-0.1 <\delta \kappa_\gamma < 0.10$) 
and $-0.1 <\lambda_\gamma < 0.16$ ($-0.16 <\lambda_\gamma < 0.24$)
for the back scattered (beamstrahlung) cases.

\vskip 0.2cm
\centerline{\epsfig{file=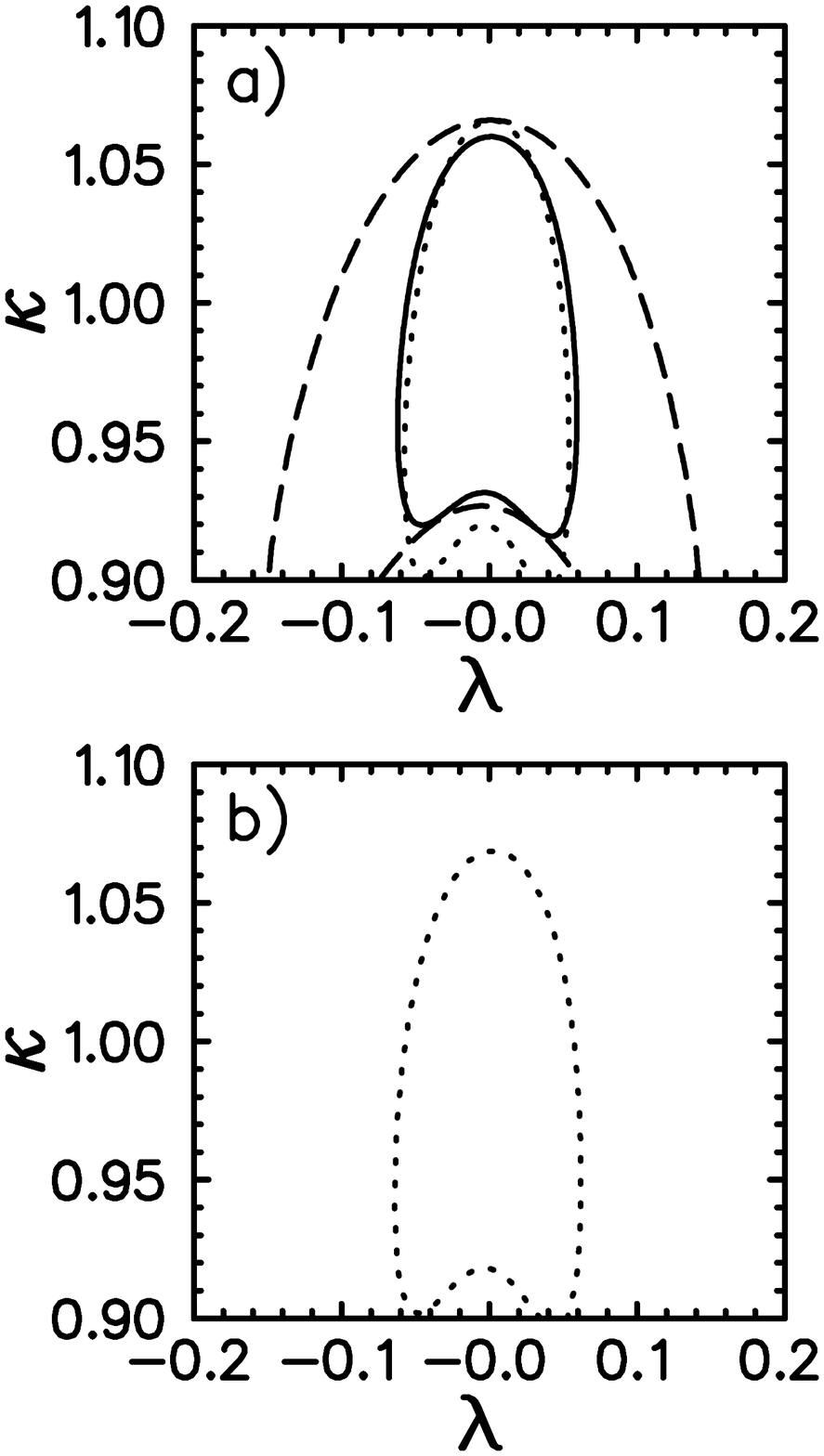,width=4.0cm,clip=}}

\noindent
{\small {\bf Figure 8:}
The achievable bounds on $\kappa_\gamma$ and $\lambda_\gamma$
at 95\% C.L. for a $\sqrt{s}=500$~GeV $e^+e^-$ collider  (a) using a
backscattered laser photon spectrum (b) using a beamstrahlung photon
spectrum.  The dashed line is based on the angular distribution
divided into four bins, the
dotted line is based on the $p_T$ distribution of the $W$ boson
divided into the four bins given in the text,
and the solid line is the combined angular and $p_T$ bounds.}

\newpage

\begin{table}
\caption{
%{\small {\bf TABLE I:} 
Bounds on $\kappa_\gamma$ and $\lambda_\gamma$ from
the processes $e^-\gamma\to \mu^-\bar{\nu}_\mu \nu_e$ and
$e^- \gamma \to W^- \to e^+ q \bar{q}$ at $\sqrt{s}=500$
using backscattered laser photon distributions (laser) and beamstrahlung
photon distributions (beam). }
\begin{tabular}{llllllll}
	&	& \multicolumn{3}{c}{$\delta\kappa_\gamma$}
	& \multicolumn{3}{c}{$\delta\lambda_\gamma$} \\
\tableline
	&	& 68\% C.L. & 90 \% C.L. & 95 \% C.L.
	& 68\% C.L. & 90 \% C.L. & 95 \% C.L. \\
\tableline
	&	&\multicolumn{6}{c}{$\sqrt{s}=500$ GeV} \\
\tableline
$\sigma_W$
	& laser
	& $^{+0.05}_{-0.05}$ & $^{+0.08}_{-0.08}$ & $^{+0.09}_{-0.10}$
	& $^{+0.14}_{-0.15}$ & $^{+0.18}_{-0.19}$ & $^{+0.20}_{-0.21}$ \\
	& beam
	& $^{+0.04}_{-0.04}$ & $^{+0.07}_{-0.07}$ & $^{+0.08}_{-0.09}$
	& \multicolumn{3}{c}{weak bounds}  \\
$\sigma_\mu$
	& laser
	& $^{+0.05}_{-0.05}$ & $^{+0.08}_{-0.08}$ & $^{+0.09}_{-0.10}$
	& $^{+0.15}_{-0.15}$ & $^{+0.19}_{-0.20}$ & $^{+0.21}_{-0.21}$ \\
	& beam
	& $^{+0.04}_{-0.05}$ & $^{+0.07}_{-0.08}$ & $^{+0.09}_{-0.09}$
	& \multicolumn{3}{c}{weak bounds}  \\
$\cos\theta_W$
	& laser
	& $^{+0.05}_{-0.05}$ & $^{+0.06}_{-0.07}$ & $^{+0.07}_{-0.07}$
	& $^{+0.09}_{-0.09}$ & $^{+0.10}_{-0.10}$ & $^{+0.10}_{-0.11}$ \\
$p_{T_W}$
	& laser
	& $^{+0.05}_{-0.05}$ & $^{+0.06}_{-0.07}$ & $^{+0.07}_{-0.08}$
	& $^{+0.04}_{-0.04}$ & $^{+0.05}_{-0.05}$ & $^{+0.05}_{-0.05}$ \\
	& beam
	& $^{+0.05}_{-0.06}$ & $^{+0.06}_{-0.07}$ & $^{+0.07}_{-0.08}$
	& $^{+0.05}_{-0.05}$ & $^{+0.05}_{-0.05}$ & $^{+0.06}_{-0.06}$ \\
%\tableline
\end{tabular}
\end{table}

\vskip 0.3cm

Unlike the case at lower energies, the limits from the muon
mode and reconstructed $W$ mode are comparable.  This is due to two
reasons: First, while the
$W$ mode is restricted to the small portion of the phase space at the
$W$ mass.  In contrast, the muon mode reflects the entire
kinematically allowed region, in particular the highest energy region
where deviations from the standard model are most pronounced.
Although the diagram we are interested in does not dominate in this
higher energy region, the interference between it and the non-resonant
diagrams are important.  Once again, this underlines the importance of
considering all contributions to the process that will actually be
observed and only then impose constraints.  Furthermore, because of the
large expected integrated luminosities the errors are dominated by
systematic errors so that the differences between the hadronic and
leptonic branching fractions become unimportant.

We repeat the above exercise for $\sqrt{s}=1$ TeV using an integrated
luminosity of  200~fb$^{-1}$.  As was the case at 500 GeV
we find that the angular  and  $p_{T_W}$  distributions give the best
constraints on anomalous couplings.  These distributions are shown in
Fig. 9 and 10 and are seen to be qualitatively similar to the 500 GeV
distributions.  The invariant mass distributions of the $q\bar{q}$ pair
is very similar to the 500 GeV except that it extends out about a factor
of two further.  To extract bounds from these distributions
we divided the angular distributions into four equal bins and the
$p_{T_W}$ distribution into the four bins;
$0-200$~GeV, 200-300~GeV, 300-400~GeV, and 400-500~GeV.
At 1 TeV there is little improvement on the
sensitivity to $\kappa_\gamma$ and about a factor of two improvement
on the sensitivity to $\lambda_\gamma$ resulting in a possible
limit of  $\delta\lambda_\gamma =\pm 0.016$ at 95\% C.L. using
the laser mode.
This measurement improvement on $\lambda_\gamma$ reflects the
greater sensitivity at higher energy.
Confidence level curves are shown in Fig. 11 with some of the
results summarized in Table II.
For $M_{q\bar{q}}>600$~GeV we obtain the 95\% C.L. sensitivity of 
$|\lambda_\gamma| < 0.03$ ($|\lambda_\gamma| < 0.05$)
for the back scattered (beamstrahlung) cases.  The $\kappa_\gamma$ 
sensitivies are relatively weak and are therefore not given.

\centerline{\epsfig{file=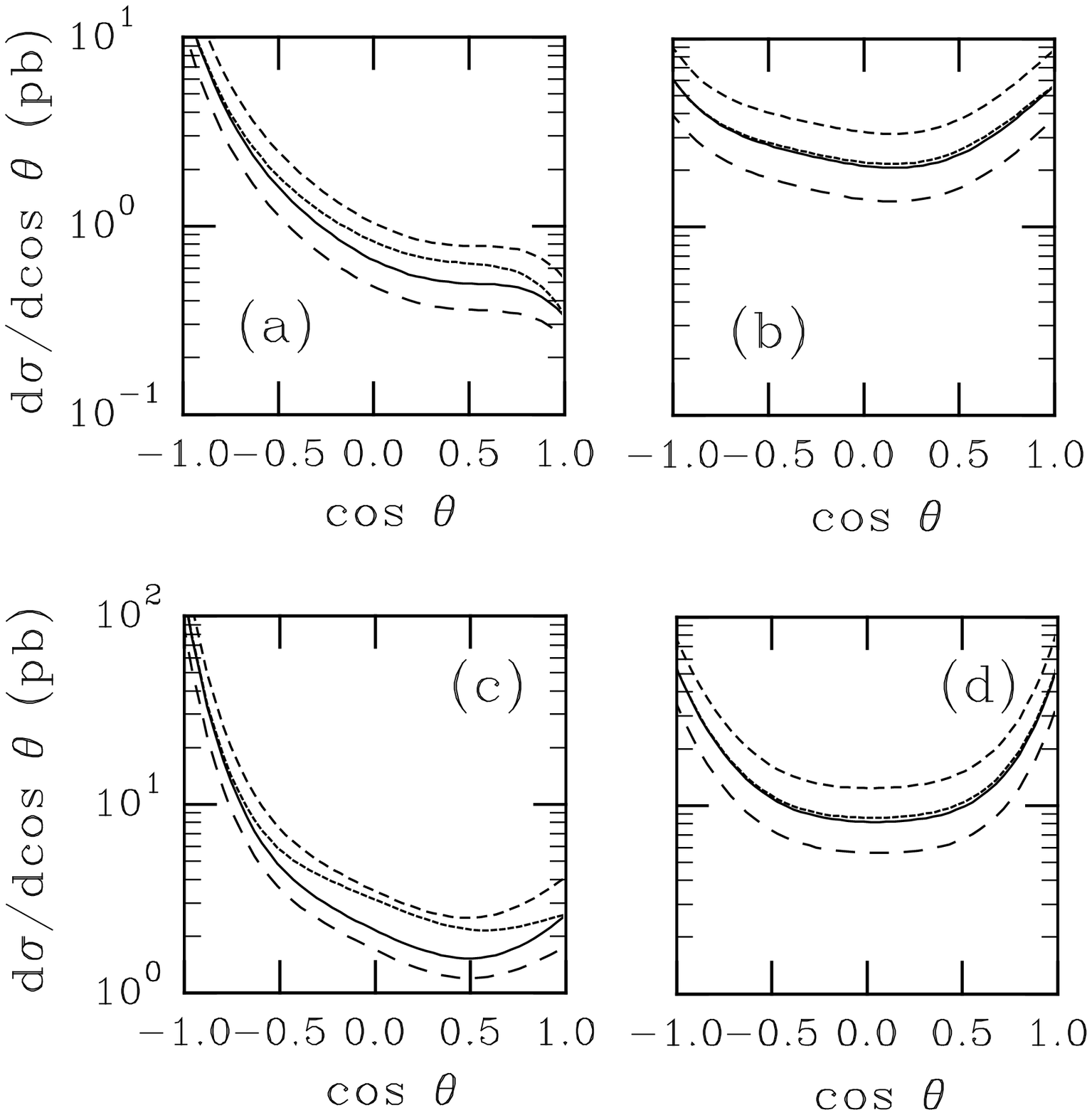,width=7.0cm,clip=}}

\noindent
{\small {\bf Figure 9:}
The angular distributions of the outgoing muon and
reconstructed $W$ boson
relative to the incoming electron for $\sqrt{s}=1$~TeV. (a) For a
muon with the backscattered laser photon spectrum, (b) for a muon
with the beamstrahlung photon spectrum, (c) for a reconstructed
$W$ boson with the backscattered laser photon spectrum, and (d) for a
$W$ boson with the beamstrahlung photon spectrum.  In all cases
the solid line is the standard model
prediction, the long-dashed line is for $\kappa_\gamma=0.6$,
$\lambda_\gamma=0$,  the short-dashed line is for $\kappa_\gamma=1.4$,
$\lambda_\gamma=0$,  and the dotted line is for $\kappa_\gamma=1$,
$\lambda_\gamma=0.1$.}

\centerline{\epsfig{file=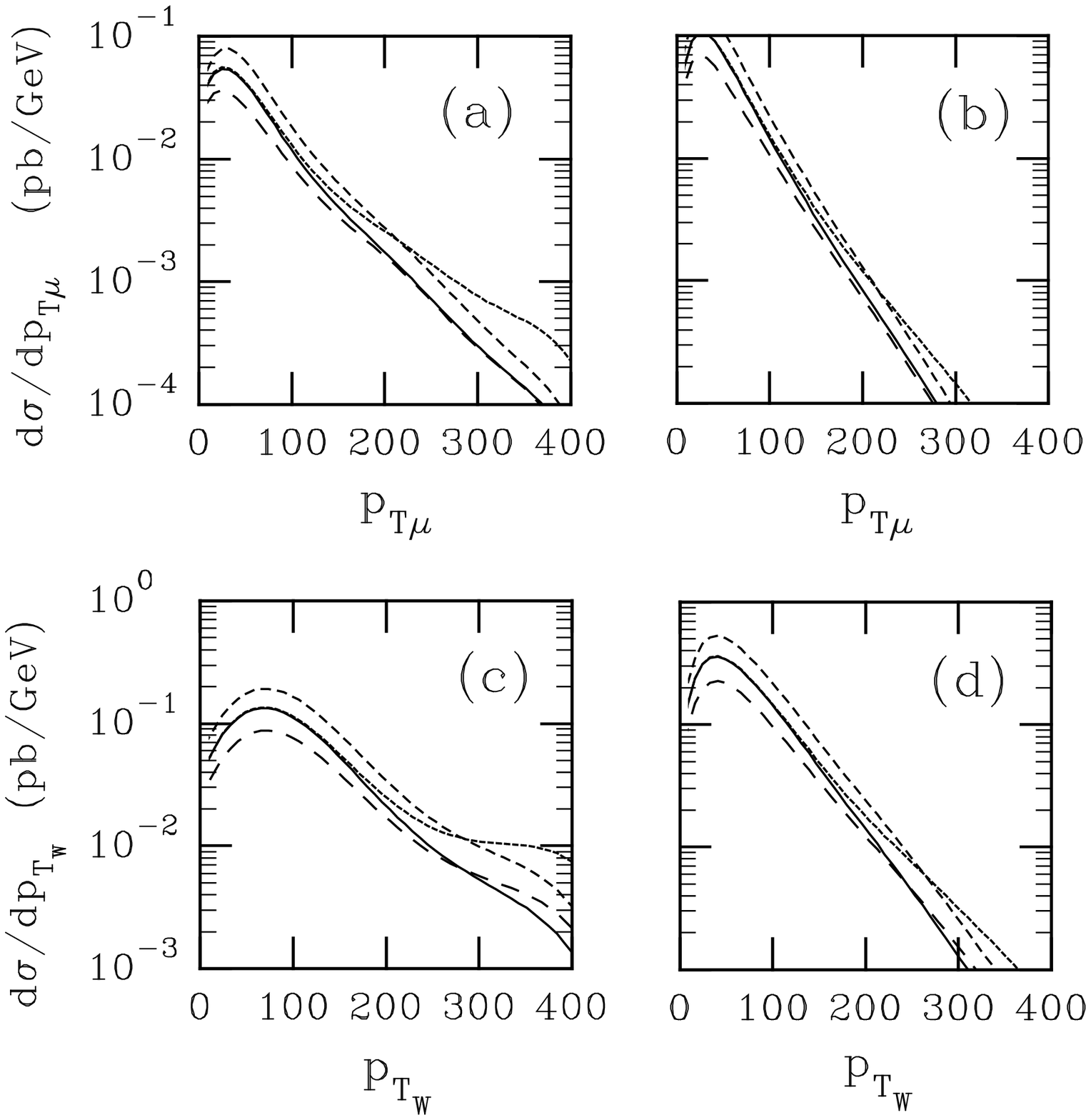,width=7.0cm,clip=}}

\noindent
{\small {\bf Figure 10:}
The $p_T$ distributions of the outgoing muon and reconstructed
$W$ boson for $\sqrt{s}=1$~TeV with the same labelling as Fig. 9.}

\centerline{\epsfig{file=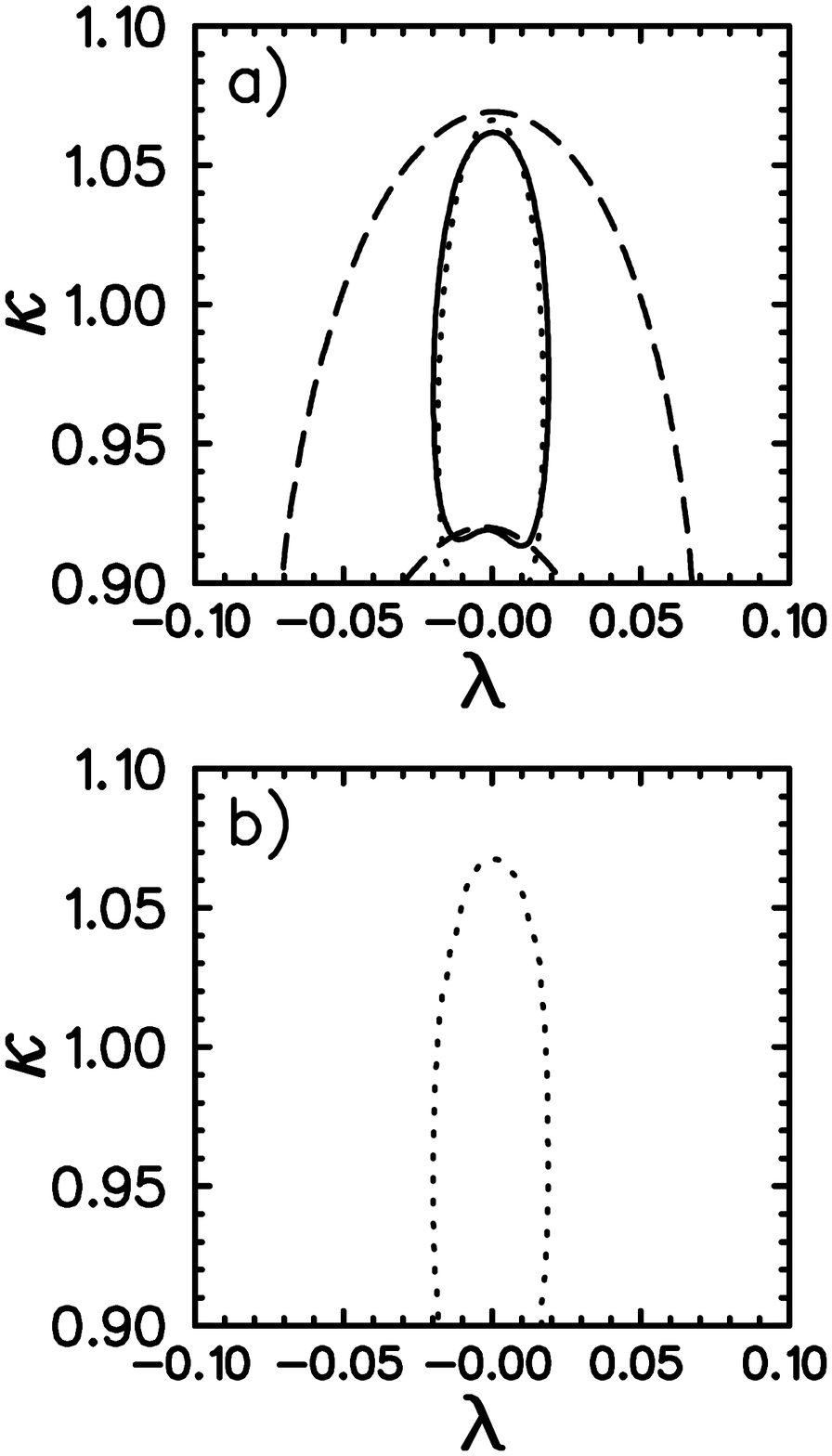,width=4.0cm,clip=}}

\noindent
{\small {\bf Figure 11:}
The achievable bounds on $\kappa_\gamma$ and $\lambda_\gamma$
at 95\% C.L. for a $\sqrt{s}=1$~TeV $e^+e^-$ collider with the same line
definitions as Fig. 8.}

%\newpage
\vskip 0.3cm
\begin{table}
\caption{
%{\small {\bf TABLE II:} 
Bounds on $\kappa_\gamma$ and $\lambda_\gamma$ from
the processes  $e^-\gamma\to \mu^-\bar{\nu}_\mu \nu_e$ and
 $e^- \gamma \to W^- \to e^+ q \bar{q}$ at 1 TeV
using backscattered laser photon distributions (laser) and beamstrahlung
photon distributions (beam).  }
\begin{tabular}{llllllll}
	&	& \multicolumn{3}{c}{$\delta\kappa_\gamma$}
	& \multicolumn{3}{c}{$\delta\lambda_\gamma$} \\
\tableline
	&	& 68\% C.L. & 90 \% C.L. & 95 \% C.L.
	& 68\% C.L. & 90 \% C.L. & 95 \% C.L. \\
\tableline
	&	&\multicolumn{6}{c}{$\sqrt{s}=1$ TeV} \\
\tableline
$\sigma_W$
	& laser
	& $^{+0.05}_{-0.05}$ & $^{+0.08}_{-0.09}$ & $^{+0.09}_{-0.10}$
	& $^{+0.07}_{-0.08}$ & $^{+0.09}_{-0.10}$ & $^{+0.10}_{-0.11}$ \\
	& beam
	& $^{+0.05}_{-0.05}$ & $^{+0.08}_{-0.08}$ & $^{+0.09}_{-0.10}$
	& \multicolumn{3}{c}{weak bounds}  \\
$\sigma_\mu$
	& laser
	& $^{+0.05}_{-0.05}$ & $^{+0.08}_{-0.09}$ & $^{+0.09}_{-0.10}$
	& $^{+0.08}_{-0.08}$ & $^{+0.10}_{-0.10}$ & $^{+0.11}_{-0.12}$ \\
	& beam
	& $^{+0.05}_{-0.05}$ & $^{+0.08}_{-0.08}$ & $^{+0.09}_{-0.10}$
	& \multicolumn{3}{c}{weak bounds}  \\
$\cos\theta_W$
	& laser
	& $^{+0.05}_{-0.06}$ & $^{+0.06}_{-0.07}$ & $^{+0.07}_{-0.08}$
	& $^{+0.04}_{-0.04}$ & $^{+0.05}_{-0.05}$ & $^{+0.05}_{-0.05}$ \\
$p_{T_W}$
	& laser
	& $^{+0.05}_{-0.07}$ & $^{+0.06}_{-0.10}$ & $^{+0.07}_{-0.11}$
	& $^{+0.013}_{-0.013}$ & $^{+0.015}_{-0.015}$ & $^{+0.016}_{-0.016}$ \\
	& beam
	& $^{+0.05}_{-0.07}$ & $^{+0.06}_{-0.10}$ & $^{+0.07}_{-0.12}$
	& $^{+0.015}_{-0.015}$ & $^{+0.017}_{-0.017}$ & $^{+0.018}_{-0.018}$ \\
\end{tabular}
\end{table}

To obtain these sensitivities to anomalous couplings we made a number
of assumptions on  $\Lambda$, the scale of new physics used in the form
factors, the beamstrahlung spectrum, and the systematic error.  We
discuss the effects of varying  these parameters starting
with the systematic error which is relevant to the 500 GeV and 1 TeV
cases. Reducing the systematic error from 5\% to 2\%
reduces the precision on $\kappa_\gamma$ roughly proportionately with
the systematic error, i.e.  a factor of two reduction in the systematic
error will tighten the limits on $\kappa_\gamma$ by roughly a factor of
two.  In contrast, the attainable constraints on $\lambda_\gamma$ does
not in general improve as much, especially for constraints obtained from the
$p_{T_W}$ distributions which give the tightest of all bounds on
$\lambda_\gamma$.  This is more pronounced  for the 500 GeV case than
the TeV case. We can see the reason for this by refering
to figures 5, 6, 9, and 10.    Varying $\kappa_\gamma$ results in an
overall shift in the cross section effecting all regions of phase space
while in contrast, the effect of $\lambda_\gamma$ grows larger with
increasing $p_{T_W}$.  The largest effect is at highest $p_T$ where the
cross section, and hence the statistics are lowest.  Thus,  at least for
the 500 GeV case, statistical errors still play a role and it indicates
that at a TeV they could also be important if the NLC does not achieve
the large integrated luminosities we have assumed.

We next consider the effect of using the beamstrahlung spectrum arising
from the G=1 beam geometry.  The only place this change has any effect
on our results is a slight improvement  on the bounds obtained using the
$p_T$ distributions.  This can be attributed to the effect the harder
photon spectrum has on the $p_T$ distribution.

Finally, we consider the effect of varying $\Lambda$, the energy scale
used in the form factors to preserve unitarity at high energy.  We took
$\Lambda=1$~TeV to obtain our results.  Changing $\Lambda$ to 500
GeV and to 2 TeV has no effect whatsoever on our
$\sqrt{s}=200$~GeV bounds.  For the 500~GeV case there is a
small decrease in the sensitivity (of the order of a few percent)
if we decrease $\Lambda$ to 500 GeV
and virtually no change when we increase it to 2 TeV.  This is true for
all measurements except those involving the $p_T$ distribution.  This is
not unexpected.  Since the form factor was introduced to suppress
anomalous couplings at high energy, if the cutoff scale is reduced it
is doing what is was introduced to do.  If this scale is much larger than the
characteristic energy scale of the process being studied, increasing it
further should have no effect.  The exception of the $p_T$ distributions
reflects that these particular observables are most sensitive to
deviations at high energy.  We find similar effects for the $\sqrt{s}=1$
TeV collider although here the changes are more pronounced.  To
demonstrate the effect on the $p_T$ distribution we plot in Fig. 12 the
$p_T$ distribution of the outgoing muon for the three values of
$\Lambda$ with $\lambda_\gamma=0.1$ using the backscattered laser
spectrum with $\sqrt{s}=1$ TeV.  We do not consider the choice of this
scale an important one.  If it were small enough to make a difference
in these measurements we would expect the new
physics to manifest itself in other ways, otherwise, the scale would be
large enough not to matter.

\vskip 0.3cm
\centerline{\epsfig{file=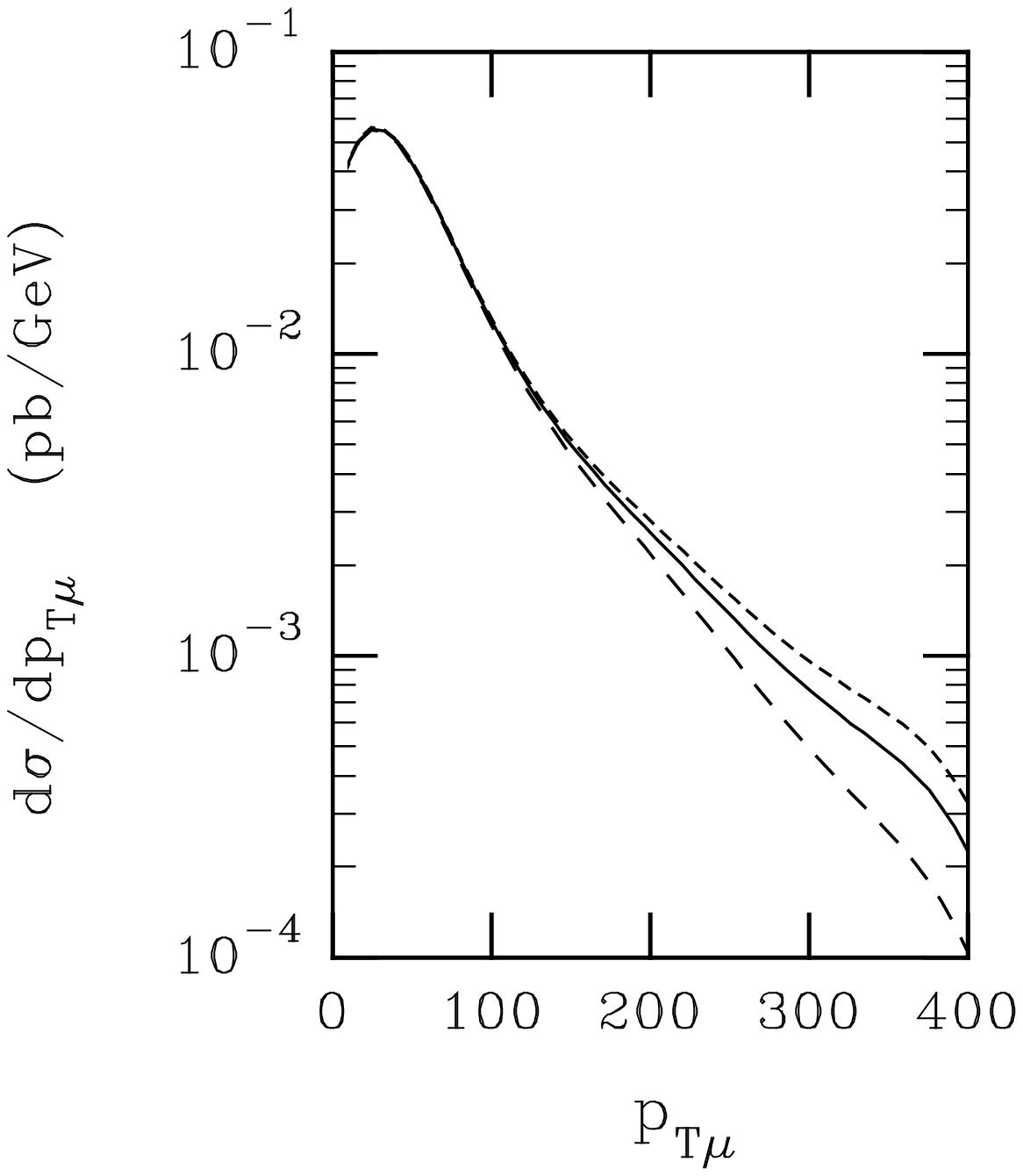,width=5.0cm,clip=}}

\noindent
{\small {\bf Figure 12:}
The $p_T$ distribution for the $\mu$ at $\sqrt{s}=1$~TeV and 
taking $\lambda_\gamma=0.1$ for different values of
$\Lambda$, the energy scale in the anomalous couplings form factor.
The short dashed line is for $\Lambda=500$~GeV, the solid line for
$\Lambda=1$~TeV, and the dashed line for $\Lambda=2$~TeV.}

\section{CONCLUSIONS}
\label{sec:conclusions}

We examined single $W$ production in $e\gamma$ collisions for a number
of collider energies and sources of energetic photons.  In our studies
we included the $W$ boson decays to final state fermions and other
processes which contribute to the same final state.
 At high energy,  the off resonance results are important since
interference effects between these other diagrams and the $W$ production
diagrams enhance the significance of  anomalous couplings, particularly
$\lambda_\gamma$.  Although these effects contribute relatively little
to the total cross section, their significance in constraining the
anomalous couplings can be large, especially at high energies and high
luminosities where these effects are statistically significant.

We found that using single $W$ production in $e^+e^-$ collisions at
$\sqrt{s}=200$~GeV in the effective photon approximation
$\kappa_\gamma$ could be measured to $\sim \pm 0.15$ and
$\lambda_\gamma$ to $\sim \pm 0.6$ at 95\% C.L..
The former measurement is
comparable to what can be achieved in $W$ pair production while the
latter is not quite as precise.  What makes this process interesting
is that it offers a means of measuring the $WW\gamma$ couplings
independently of the $WWZ$ couplings which is far from trivial in the
$W$ pair production process.  

For the high energy, high luminosity, NLC $e^+e^-$ collider we
considered both a backscattered laser photon spectrum and a
beamstrahlung photon spectrum.  For $\sqrt{s}=500$~GeV
$\delta\kappa_\gamma \simeq \pm 0.07$ and $\delta\lambda_\gamma\simeq
\pm 0.05$
and for $\sqrt{s}=1$~TeV,
$\delta\kappa_\gamma \simeq \pm 0.07$ and $\delta\lambda_\gamma\simeq
\pm 0.016$ using the backscattered laser approach.  Using the
beamstrahlung photon spectrum is only slightly less sensitive.
The measurement of $\kappa_\gamma$
 is approaching the level of radiative
corrections and might be sensitive to new physics at the loop level.
On the other hand, it is expected that the sensitivity to $\lambda$
would have to be at least an order of magnitude more sensitive to be
interesting.  From our analysis it does not appear that there is any
overwhelming advantage to go to higher energies to study the trilinear
gauge boson couplings using $e\gamma$ collisions.

\acknowledgments

The authors are most grateful to Tim Barklow, Pat Kalyniak, Dean Karlen,
Francis Halzen
and Paul Madsen for
helpful conversations and to Concha Gonzalez-Garcia for helpful
communications.
This research was supported in part by the Natural Sciences and Engineering
Research Council of Canada.  MAD is grateful for the hospitality of the 
Physics Department at Carleton University where most of this anaylsis was 
performed.

\appendix{Helicity Amplitudes}

In this appendix, we summarize  the CALKUL
spinor technique \cite{calkul} and give the helicity amplitudes for the
process $e\gamma \to \nu f \bar{f}'$ where $f\bar{f}'$ can be
$\mu \bar{\nu}_\mu$ or $q\bar{q}'$. We do not
go into any detail and refer the interested reader
to the literature and references therein.
We limit our discussion to
massless fermions and massless external gauge bosons,
which apply to our problem.
The propagators for the fermions and gauge bosons have the same form as in
the usual trace technique so  we will not discuss them here.

The spinor technique results in reducing strings of spinors
and gamma matrices to sandwiches of spinors which can be evaluated easily.
In doing so, one makes extensive use of the {\it right} and {\it left}
projection operators defined by
$ \omega_{\pm} = {1\over2}(1{\pm}\gamma_5 ) $.
One defines two four-vectors, $k_0^{\mu}$ and $k_1^{\mu}$, which obey the
following relations:
$$k_0\cdot k_0 = 0, \phantom{ssss}k_1\cdot k_1 = -1,\phantom{ssss}k_0\cdot
k_1 = 0. $$
and the basic spinors:
$$ u_{-}(k_0){\bar u}_{-}(k_0) = {\omega}_{-}{\not k}_0 $$
and
$$ u_{+}(k_0) = {\not k}_1 u_{-}(k_0) $$
Note that in the massless limit, one can use $u$ and ${\bar u}$ to describe
both particles and antiparticles, with the spin sum
$\sum_{\lambda} u_\lambda (p) {\bar u}_\lambda (p) = {\not p} $.
These two spinors are the building blocks for any spinor of lightlike
momentum p :
$$ u_{\lambda}(p)= { { {\not p}u_{-\lambda}(k_0)} \over
{\sqrt{2\;p\cdot k_0} } }$$
Two identities are essential for the reduction of the strings; the spin sum
given above and
the Chisholm identity:
$$ {\bar u}_\lambda (p_1){\gamma}^{\mu} u_\lambda (p_2) {\gamma}_{\mu}
   \equiv 2 u_\lambda (p_2) {\bar u}_\lambda (p_1) +
          2 u_{-\lambda}(p_1) {\bar u}_{-\lambda}(p_2)$$
where ${\lambda}$ is ${\pm 1}$ and represents the helicity state.
These two identities allow one to reduce strings of spinors and gamma
matrices to sandwiches of spinors.
Only two of the four possible sandwiches are non-zero:
$$ s(p_1,p_2)\equiv {\bar u}_+(p_1) u_-(p_2) = -s(p_2,p_1) $$
and
$$ t(p_1,p_2)\equiv {\bar u}_-(p_1) u_+(p_2) = s(p_2,p_1)^{*}. $$
Once the amplitude has been reduced to a series of factors of $s( p_i ,p_j )$
and $t( p_k,p_l )$, the expressions can be
evaluated
by computer. A judicious choice of the four-vectors $k_0^{\mu}$ and
$k_1^{\mu}$ simplifies the evaluation of the $s$ and $t$ terms. For our
calculation, we used the definition of ref. \cite{calkul};
$$ p_i^{\mu} = ( p_i^0,p_i^x,p_i^y,p_i^z ) $$
$$ k_0^{\mu} = ( 1,1,0,0 ) $$
$$ k_1^{\mu} = ( 0,0,1,0 ) $$
to obtain
$$s(p_1,p_2) = (p_1^y + i p_1^z )
{\sqrt{ p_2^0 - p_2^x}\over{\sqrt{ p_1^0 - p_1^x}}} -
( p_2^y + i p_2^z )
{\sqrt{ p_1^0 - p_1^x}\over{\sqrt{ p_2^0 - p_2^x}}} $$

These forms are ideally suited for programming. When dealing with several
diagrams, one simply evaluates the amplitudes of each diagram
as complex numbers and squares the sum of the  amplitudes in order to
obtain the $\vert amplitude \vert ^2$.

To include massless gauge bosons one represents, as usual,
the gauge boson by its
polarization vector. Following Kleiss and Sterling
we use the definition:
$$ \epsilon_\lambda^{\mu}(k) \equiv {1\over{ \sqrt{4\; p\cdot k}}}
{\bar u}_\lambda (k) {\gamma}^\mu u_\lambda (p)  $$
where $p^\mu$ is any lightlike four-vector not collinear to $k^\mu$ or
$k_0^\mu$. The choice of $p^\mu$ acts as a choice of gauge
and provides a powerful verification
of gauge invariance; it can be shown that two different choices of
$p^\mu$ will lead to two expressions that will differ by a term proportional
to the photon momentum. When
dotted into the amplitude, this extra term must vanish identically because
of gauge invariance. Hence, two different choices of $p^\mu$ must give
exactly the same answer. If they don't, there is a mistake in the amplitude.
Generally, we choose $p^\mu$ to be one of the four-vectors of the problem
at hand.

Using this technique we obtain for the helicity amplitudes
corresponding to the Feynman diagrams of Fig. 2, using the notation
$M=ieg^2 \tilde{M} /\sqrt{4p_\gamma \cdot k}$:

\begin{eqnarray}
\tilde{M}_{LL}^a & = & -{ 2 \over {s \;  P_W(q+\bar{q}) }}
 t(p_\nu , q ) \; s(\bar{q}, p_e ) \; t(p_e, p_\gamma )\; s(k, p_e )\\
\tilde{M}_{LR}^a & = & -{ 2 \over {s   P_W(q+\bar{q}) }}
 t(p_\nu, q ) \; s(p_\gamma , p_e )
\left[ s(\bar{q}, p_e)\; t(p_e , k )+ s(\bar{q},p_\gamma )\; t(p_\gamma, k)
\right]\\
\tilde{M}_{LL}^b & = & {1\over{ P_W(p_e-p_\nu)\; P_W(q+\bar{q})}} \nonumber\\
& &  \times \Big\{
-2 t( p_{\nu}  ,q)\; s(\bar{q}, p_e)\;
[t(p_\gamma, p_e)\; s(p_e,k ) -t(p_\gamma, p_\nu)\; s(p_\nu, k)]
\nonumber \\
& & \qquad
+(1 +\kappa + \hat{\lambda} (p_e-p_\nu)^2 )
t(q, p_\gamma) \; s(k,\bar{q}) \; t(p_\nu, p_\gamma) \; s(p_\gamma,
p_e) \nonumber \\
& &\qquad
-(1+\kappa+ \hat{\lambda}(q+\bar{q})^2)
t( p_\nu, p_{\gamma}) \; s(k, p_e) \; t(q,p_\gamma) \; s(p_\gamma,
\bar{q}) \nonumber \\
& & \qquad
+\hat{\lambda}
[t(p_\gamma, p_e) \; s(p_e,k) - t(p_\gamma, p_\nu) \; s(p_\nu, k)]
\nonumber \\
& & \qquad \qquad \times
t(q, p_\gamma) \; s(p_\gamma, \bar{q}) \;
t(p_\nu, p_\gamma) \; s(p_\gamma, p_e)
\Big\} \\
\tilde{M}_{LR}^b & = & {1\over{ P_W (p_e-p_\nu)\; P_W(q+\bar{q} ) } }
\nonumber\\
& &  \times \Big\{
-2 t( p_{\nu}  ,q)\; s(\bar{q}, p_e)\;
[s(p_\gamma,  p_e)\; t(p_e,k ) -s(p_\gamma, p_\nu)\; t(p_\nu, k) ]
\nonumber\\
& & \qquad
+(1 +\kappa + \hat{\lambda} (p_e-p_\nu)^2 )
t(q, k) \; s(p_\gamma,\bar{q}) \; t(p_\nu, p_\gamma) \; s(p_\gamma,
p_e) \nonumber\\
& &\qquad
-(1+\kappa+ \hat{\lambda}(q+\bar{q})^2)
t( p_\nu, k) \; s(p_\gamma, p_e) \; t(q,p_\gamma) \; s(p_\gamma,
\bar{q}) \nonumber \\
& & \qquad
+\hat{\lambda}
[s(p_\gamma, p_e) \; t(p_e,k) - s(p_\gamma, p_\nu ) \; t(p_\nu, k)]
\nonumber\\
& & \qquad \qquad \times
t(q, p_\gamma) \; s(p_\gamma,\bar{q}) \;
t(p_\nu, p_\gamma) \; s(p_\gamma, p_e)
\Big\} \\
\tilde{M}_{LL}^c & = & +{ {2  Q_{\bar{f } } } \over { (p_\gamma-\bar{q})^2 \;
P_W(p_e-p_\nu)}} \nonumber \\
& & \qquad \qquad \times t( q, p_{\nu} ) \; s( k, \bar{q})
[ s(p_e, q)\;t( q, p_{\gamma} ) +s(p_e, p_{\nu} ) \; t( p_\nu, p_{\gamma} )]\\
\tilde{M}_{LR}^c & = &  +{ {2  Q_{\bar{f}} } \over { (p_\gamma-\bar{q})^2 \;
P_W(p_e-p_\nu)}} \nonumber \\
& & \qquad\qquad\times t( q,p_\nu) \; s( p_{\gamma}, \bar{q} )
[s(p_e, q )\; t(q,k) + s(p_e, p_\nu)\; t(p_\nu, k ) ] \\
\tilde{M}_{LL}^d & = &
+{ {2  Q_f } \over{ (q- p_\gamma)^2\; P_W(p_e-p_\nu)} } \nonumber \\
& & \qquad \qquad \times t( q, p_\gamma ) \; s( p_e, \bar{q})
[ s(k,q )\;t( q, p_\nu ) -s(k,p_\gamma ) \; t( p_\gamma, p_\nu ) ] \\
\tilde{M}_{LR}^d & = &
+{ {2  Q_f } \over{ (q- p_\gamma)^2 \; P_W(p_e-p_\nu) } }
t( q,k) \; s( p_e, \bar{q} )  \; s(p_\gamma, q )\; t(q, p_\nu)
\end{eqnarray}
where the propagators are defined by
\begin{equation}
P_W(p) = [ p^2 -M_W^2 +i \Gamma_W M_W ] .
\end{equation}
The first subscript of the amplitudes refers to the helicity of the
electron and the second subscript to the helicity of the photon.
The amplitudes correspond to the diagrams of Fig.  2 where the
four momenta $p_e, p_\gamma, p_\nu, q$ and $\bar{q}$ are
defined.   $Q_f=Q_d=-1/3$ and $Q_{\bar{f}}=Q_u=+2/3$.
Helicity amplitudes not explicitly written down are zero.  To
obtain the cross section the amplitudes for given electron
and photon helicities are summed over and squared.  These are then
averaged to obtain the spin averaged matrix element squared and
finally integrated over the final state phase space to yield the cross section.

\appendix{The Photon Distributions}

To obtain the cross sections in the main text we convolute the
$e\gamma$ cross section with the relevant photon distributions:
\begin{equation}
\sigma = \int_0^1 \; f_{\gamma/e} (x) \; \sigma(e\gamma\to W \nu) \; dx
\end{equation}
where the various photon distributions, $f_{\gamma/e}(x)$, are given
below.

\subsection{Back-Scattered Laser Photons}

Intense high energy photon beams can be obtained by backscattering a low
energy laser off of a high energy electron beam.  The energy spectrum of
the back-scattered laser photons is given by \cite{laser}
\begin{equation}
f_{\gamma/e}^{laser}(x,\xi) = {1\over{D(\xi)}} \left[ { 1 -x
+{1\over{1-x}} - {{4x}\over {\xi (1-x)}} + {{4x^2}\over{\xi^2(1-x)^2}}
}\right]
\end{equation}
where the fraction $x$ represents the ratio of the scattered photon
energy, $\omega$, and the initial electron energy, $E$,  ($x=\omega/E$) and
\begin{equation}
D(\xi) = \left( { 1- {4\over \xi} - {8 \over {\xi^2}} } \right)
\ln(1+\xi) + {1\over 2} + {8\over\xi} - {1\over{2(1+\xi)^2}}
\end{equation}
with
\begin{equation}
\xi= {{4E\omega_0}\over {m^2_e}} \cos^2 {\alpha_0\over 2} \simeq
{{2\sqrt{s}\omega_0}\over {m^2_e}}
\end{equation}
and $\omega_0$ is the laser photon energy and $\alpha_0 \sim 0$ is the
electron-laser collision angle.  The maximum value of $x$ is
\begin{equation}
x_m ={\omega_m \over E} = {\xi \over {(1 + \xi)}}.
\end{equation}
Because of the onset of $e^+e^-$ pair production between backscattered
and laser photons, conversion efficiency drops considerably for
$x>2+2\sqrt{2} \approx 4.82$.  We use this value which for 250 GeV
electrons corresponds to a laser energy of about 1.26 eV.

The photon spectrum is sensitive to the the product $\lambda_e
P_\gamma$ where $\lambda_e$ is the mean electron helicity and
$P_\gamma$ is the mean laser helicity.  Larger values of $\lambda_e
P_\gamma$ give a harder more monochromatic photon spectrum. Measuring
the actual $\lambda_e$ introduces systematic errors so we assume that
the electron beam is unpolarized.  In constrast the laser can be
easily polarized almost completely.  The amount of polarization is
energy dependent. Assuming $P_\gamma=1$ the average
helicity $\xi_2$ of the photon beam is given by
\begin{equation}
\xi_2= - {{\xi (\xi - 2 x - \xi x )(2-2 x +x^2)}\over{2\xi^2-4 x\xi -4
\xi^2 x + 4x^2 + 4 \xi x^2 + 3 \xi^2 x^2 - \xi^2 x^3}}
\end{equation}
The long dashed line in Fig. 13 shows the spectrum of photons for an
unpolarized
laser and the medium dashed line shows the photon spectrum with helicity
$-P_\gamma$. Note that the most energetic photons are always polarized
with opposite helicity to that of the laser photons.

\subsection{Beamstrahlung Photons}

The interpenetration of the dense electron and positron bunches in
future $e^+e^-$ colliders generates strong accelerations on the
electrons and positrons near the interaction point which gives rise to
synchrotron radiation which is referred to in the literature as
beamstrahlung \cite{beam1,beam2,beam3}.  Beamstrahlung depends
strongly on machine parameters such as luminosity, pulse rate, and
bunch geometry.
The distribution function of beamstrahlung photons can
be written in the following approximation:
\begin{equation}
f_{\gamma/e}^{beam}(x,b) = f_{\gamma/e}^{(-)}(x,b)\;\Theta(x_c-x) +
f_{\gamma/e}^{(+)}(x,b)\;\Theta(x-x_c)
\end{equation}
where, as before, $x$ is the fraction of the beam energy carried by
the photon, $b$ is the impact parameter of the produced $\gamma$, and
$x_c$ separates low and high photon energy regions where the different
approximations to $f_{\gamma/e}^{beam}(x,b)$ are used.  The
distribution used for small and intermediate values of $x$ is given by
\begin{equation}
f_{\gamma/e}^{(-)}(x,b) \simeq {{CK}\over {\Upsilon^{1/3}}}
\left[ {{1+(1-x)^2}\over{x^{2/3}(1-x)^{1/3}} } \right]
\left\{ { 1 + {1\over{6 C \Upsilon^{2/3}}}
\left(  {x \over {1-x}} \right)^{2/3}
\exp \left[ { { 2\over {3\Upsilon}} {x\over {(1-x)}} } \right]
}\right\} ^{-1}
\end{equation}
where $C=-Ai'(0)=0.2588$, and $Ai(x)$ is Airy's function.
For large values of $x$ the distribution is given by
\begin{equation}
f_{\gamma/e}^{(+)}(x,b)\simeq {K \over {2\sqrt{\pi} \Upsilon^{1/2}}}
\left[ { { 1-x(1-x)}\over { x^{1/2} (1-x)^{1/2} } } \right]
\exp \left[ { - {2\over {3 \Upsilon} } {x \over {(1-x)} } } \right].
\end{equation}
The value $x_c$ is such that $f^{beam}_{\gamma/e}$ is continuous at
$x=x_c$.  It  depends on the machine design; for the 500~GeV NLC, $x_c \simeq
0.48$.  The dimensionless quantities $K$ and $\Upsilon$ are defined as
\begin{eqnarray}
K & = & 2\sqrt{3} \alpha {{\sigma_z E_\perp}\over m } \nonumber \\
\Upsilon & = & {{p E_\perp}\over {m^3}}
\end{eqnarray}
where $m$ and $p$ are the electron mass and momentum, and $E_\perp$ is
the transverse electric field inside a uniform elliptical bunch of
dimensions $l_{x,y} =2\sigma_{x,y}$ and $l_z =2 \sqrt{3} \sigma_z$
\begin{equation}
E_\perp = {{ N\alpha}\over {2\sqrt{3} (\sigma_x + \sigma_y) \sigma_z}}
\left( { { {b_x^2}\over {\sigma_x^2} } + { {b_y^2}\over {\sigma_y^2} }
} \right)^{1/2}
\label{eq:et}
\end{equation}
with $N$ being the number of particles in the bunch.

For the case of beamstralung we have to average over the impact
parameter in addition to integrating over the energy fraction, $x$.
Using elliptical coordinates the expression in the brackets of
Eq.~\ref{eq:et} becomes
\begin{equation}
\left( { { {b_x^2}\over {\sigma_x^2} } + { {b_y^2}\over {\sigma_y^2} }
} \right)^{1/2} =2b
\end{equation}
and
\begin{equation}
f_{\gamma/e}^{beam}(x)= \int_0^1 f_{\gamma/e}^{beam}(x,b) 2b \; db
\end{equation}

The photon luminosity of beamstrahlung is very sensitive to the
transverse shape of the beam.  The aspect ratio
\begin{equation}
G= {{\sigma_x +\sigma_y} \over {2 \sqrt{\sigma_x \sigma_y} } }
\end{equation}
provides a good measure of beamstrahlung with large photon
luminosities associated with small values of $G$.  For high photon
luminosity one tunes to round beams, $G=1$.  For the original NLC
design $G\simeq 2.7$.  We include the beam parameters for a number of
NLC options in Table III.

\vskip 0.2cm
\centerline{\epsfig{file=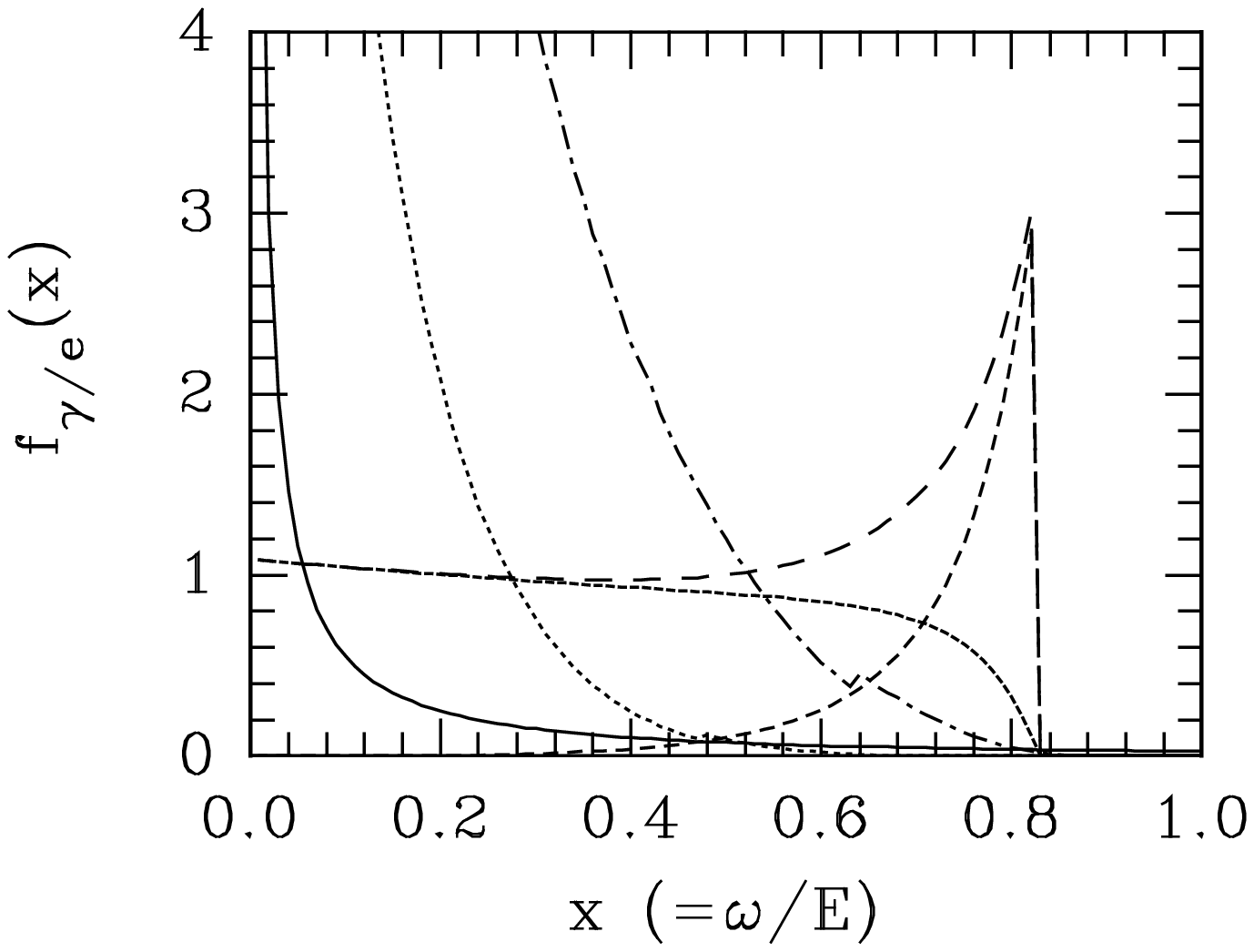,width=6.0cm,clip=}}

\noindent
{\small {\bf Figure 13:}
The photon distributions described in the text.  The solid
line  is the Weizacker-Williams distribution, the long dashed line is
the backscattered photon distribution, the medium dashed line is the
backscattered photon distribution for photons with opposite polarization
to that of the laser,
the short dashed line is the backscattered photon distribution for
photons with the same polarization as the laser,
the dotted line is the beamstrahlung
spectrum for $\sqrt{s}=500$ GeV with G=2.7, and the dash-dot line is
with G=1.}

\newpage
\begin{table}
\caption{
%{\small {\bf TABLE III:} 
NLC Machine Parameters}
\begin{tabular}{lcccc}
	& \multicolumn{2}{c}{NLC} & \multicolumn{2}{c}{TLC} \\
\tableline
$E_{cm} $ (TeV)	& 0.5 & 0.5 & 1 & 1 \\
$L$ (cm$^{-2}$sec$^{-1}$) & $ 9\times 10^{33}$ & & & \\
$N$	& $1.67\times 10^{10}$ & $1.67\times 10^{10}$
	& $1.67\times 10^{10}$ & $1.67\times 10^{10}$ \\
$\sigma_z$ (cm) & 0.011 & 0.011 & 0.011 & 0.011 \\
$\sigma_y$ (cm) & $6.5 \times 10^{-7}$ & $3.3 \times 10^{-6}$
	& $1.7 \times 10^{-5}$ & $3.3 \times 10^{-6}$ \\
$\sigma_x$ (cm) & $1.7 \times 10^{-5}$ & $3.3 \times 10^{-6}$
	& $1.7 \times 10^{-5}$ & $3.3 \times 10^{-6}$\\
G & 2.7 & 1.0 & 2.7 & 1.0 \\
$x_c$	& 0.48 & 0.64 & 0.58 & 0.72 \\
\end{tabular}
\end{table}

\subsection{Classical Bremstrahlung}

Finally we consider conventional bremsstrahlung of photons by electrons
which we use for the $\sqrt{s}=200$~GeV case and which also contributes
to the photon
luminosity when we consider beamstrahlung.  We use the well-known
Weisz\"acker-Williams distribution \cite{ww} which we include for completeness:
\begin{equation}
f_{\gamma/e}^{WW}(x, E_{max}) = {\alpha \over {2\pi}}
{{1 + (1-x)^2 } \over x} \ln \left( { {E^2_{max}}\over {m_e^2} } \right)
\end{equation}
where $E_{max}$ is the electron beam energy.


\begin{references}
\bibitem{barklow}
These values are very rough estimates based on Monte
Carlo studies. T. Barklow,  private communications.
\end{references}
\end{document}